\documentclass[fleqn,usenatbib]{mnras}

\usepackage{newtxtext,newtxmath}

\usepackage[T1]{fontenc}
\usepackage[dvipsnames]{xcolor}

\DeclareRobustCommand{\VAN}[3]{#2}
\let\VANthebibliography\thebibliography
\def\thebibliography{\DeclareRobustCommand{\VAN}[3]{##3}\VANthebibliography}

\usepackage{graphicx}	
\usepackage{amsmath}	
\usepackage{multirow}
\usepackage{array}
\usepackage{hyperref}

\title[BAO scale with the SDSS blue galaxies]
{Baryon acoustic scale at $z_{\mbox{\footnotesize eff}} = 0.166$ with the SDSS blue galaxies}

\author[Avila et al.]
{Felipe Avila,$^{1}$\thanks{e-mail: felipeavila@on.br}
Edilson de Carvalho,$^{2}$ 
Armando Bernui,$^{1}$ 
Hanna Lima,$^{3}$
and Rafael C. Nunes$^{3,4}$\\ 
$^{1}$Observat\'orio Nacional, Rua General Jos\'e Cristino 77, 
          S\~ao Crist\'ov\~ao, 20921-400 Rio de Janeiro, RJ, Brazil \\
$^{2}$Centro de Estudos Superiores de Tabatinga, Universidade do Estado do Amazonas, 
69640-000, Tabatinga, AM, Brazil \\
$^{3}$ Instituto de Física, Universidade Federal do Rio Grande do Sul, 91501-970 Porto Alegre RS, Brazil \\
$^{4}$ Divis\~ao de Astrof\'isica, Instituto Nacional de Pesquisas Espaciais, Avenida dos Astronautas 1758, S\~ao Jos\'e dos Campos, 12227-010, SP, Brazil}

\date{Accepted XXX. Received YYY; in original form ZZZ}

\pubyear{2015}

\begin{document}
\label{firstpage}
\pagerange{\pageref{firstpage}--\pageref{lastpage}}
\maketitle

\begin{abstract}
The Baryon Acoustic Oscillations (BAO) phenomenon provides a unique opportunity to establish a standard ruler at any epoch in the history of the evolving universe. The key lies in identifying a suitable cosmological tracer to conduct the measurement. In this study, we focus on quantifying the sound horizon scale of BAO in the Local Universe. Our chosen cosmological tracer is a sample of blue galaxies from the SDSS survey, positioned at the effective redshift $z_{\mbox{\footnotesize eff}} = 0.166$. Utilizing Planck-CMB input values for redshift-to-distance conversion, we derive the BAO scale $s_{\scalebox{0.65}{\rm BAO}} = 100.28 ^{+10.79} _{-22.96}$ Mpc$/h$ at the 1$\sigma$ confidence level. Subsequently, we extrapolate the BAO signal scale in redshift space: $\Delta z_{\scalebox{0.6}{\rm BAO}}(z_{\rm eff}=0.166)=0.0361^{+0.00262}_ {-0.0055}$. This measurement holds the potential to discriminate among dark energy models within the Local Universe. To validate the robustness of our methodology for BAO scale measurement, we conduct three additional BAO analyses using different cosmographic approaches for distance calculation from redshifts. These tests aim to identify possible biases or systematics in our measurements of $s_{\scalebox{0.65}{\rm BAO}}$. Encouragingly, our diverse cosmographic approaches yield results in statistical agreement with the primary measurement, indicating no significant deviations. Conclusively, our study contributes with a novel determination of the BAO scale in the Local Universe, at $z_{\mbox{\footnotesize eff}} = 0.166$, achieved through the analysis of the SDSS blue galaxies cosmic tracer.
\end{abstract}




\section{Introduction}\label{sec:Introduction}

Embedded in the three-dimensional (3D) map of the observed universe there are imprints of a primordial phenomenon, the Baryon Acoustic 
Oscillations~\citep[BAO;][]{Peebles70,Sunyaev70,Eisenstein05,Cole05}. 
The BAO signature appears on scales $\sim \!100\,\text{Mpc}/h$ and to reveal it one has to survey large spatial volumes, $\sim \!1\, \text{Gpc}^3 /h^{3}$, with number density of cosmic objects around $10^{-4} ~h^{3}/{\text{Mpc}}^{3}$, 
features now achieved thanks to the efforts of current astronomical survey collaborations, like the Sloan Digital Sky Survey (SDSS), Dark Energy Survey, 6dF Galaxy Survey, and 
WiggleZ~\citep{Percival10,Beutler11,Blake11,Abbott19,Alam21}. 
The detection of the BAO signature is important, not only because it confirms primordial physical processes but mainly, because its measurement provides a reliable standard ruler and is therefore used to make accurate measurements of cosmic distances. 
In fact, through diverse cosmological tracers mapped at different epochs of the universe, one can measure the BAO signal at several redshifts to unambiguously reveal the dynamics of the 
universe~\citep{Bond84,Einsenstein98,Bassett10,Weinberg13}.

The BAO signature is weak in the galaxy-galaxy correlations, for this, it is statistically revealed in numerically dense catalogues using the 2-point correlation function (2PCF) in, at least, two ways. 
The first approach, based on three observational data: two angles for the sky angular position and redshift, needs to assume a fiducial cosmology to transform the redshift of each cosmic object into its radial distance and with the two angles one calculates the comoving distance between all possible pairs to construct the 2PCF; the BAO signal obtained with this approach determines the sound horizon scale at the end of the baryon drag epoch, $r_s \equiv s_{\scalebox{0.65}{\rm BAO}}$, and the spherically averaged distance $D_V$~\citep{Eisenstein05,Alam21,Beutler11,Blake11,Carter18}. 
The second approach uses 2D information: one analyzes objects located in a thin redshift shell, where the data used are the two angles that determine their position on the sky. 
The objects are projected on the celestial sphere, then knowing the angular coordinates 
of each cosmic object one calculates the angular distances between pairs and constructs the 
2-point angular correlation function (2PACF), where the BAO angular scale provides a 
measure of the angular diameter distance $D_{\!A}$, if $r_s$ is known. 
To minimize projection effects that would affect this measure the data should be in a thin redshift shell~\citep{Sanchez11,Carnero12,Edilson18,Edilson21}. 

As~\cite{Crocce08} have noticed, an interesting challenge of BAO 
analyses in the Local Universe, $z \ll 1$, is the measurement of the BAO signature due to non-linear clustering process that 
smoothes the acoustic peak, thus decreasing the statistical significance of the detection. This makes sense because our Local Universe corresponds to the part of the universe where gravitational attraction acted during the longest period of cosmic time, with redshift $z \simeq 0$, smearing out the primordial BAO sphere and impacting the measurement uncertainty.

This challenge, in the BAO analyses of 3D cosmic object distributions, is often circumvented by assuming a fiducial cosmology to model the 2PCF, including the effect due to non-linear processes (see, e.g.,~\cite{Beutler11, Carter18}). 
However, an alternative approach to studying BAO at low redshifts will 
be to choose a cosmological tracer with bias relative to matter 
close to 1, i.e., $b \simeq 1$, because such cosmic objects do not 
form highly clustered regions, minimizing non-linear effects at $z \ll 1$.

As a matter of fact, detailed examinations of the galaxy clustering dependence on colour and luminosity, particularly in samples of red and blue galaxies, were developed as large astronomical surveys emerged~\citep{Zehavi05,Croton07,Ross14,Mohammad18}. 
Blue galaxies are {\em late-type galaxies} with significant star formation, meaning they are unlikely to be found in 
high density regions~\citep{Gerke07}. 
This characteristic is reflected in the clustering statistics as the 2PCF, where, on small scales, red galaxies of any luminosity are more clustered than blue galaxies of any luminosity~\citep{Zehavi05}. 
In general, the clustering strength, or {\em bias}, increases 
for galaxies with greater luminosity and redder colour, being colour more predictive of the large-scale environment (more than other properties like the morphology). 
These different clustering properties are reflected in their relative bias: 
the red galaxies have a $40\%$ larger bias than the blue ones 
$b_{\text{red}} / b_{\text{blue}} 
= 1.39 \pm 0.04$~\citep{Ross14}.


%

Therefore, to reveal the 3D BAO signal and perform a low-redshift BAO-scale measurement with minimal model assumptions we choose the blue galaxies, a cosmic tracer that shows reduced effects of non-linear clustering because they are found in low density regions~\citep{Gerke07,Mohammad18}, making it possible to fit its 2PCF without assuming a cosmological model.


%

Accordingly, we shall perform 3D-BAO analyses with an ensemble 
of SDSS blue galaxies at low redshift from the Sloan Digital Sky Survey (SDSS). 
The set of blue galaxies is, indeed, a robust cosmological tracer that can be used to investigate the BAO features in the Local Universe~\citep{Ross14,Carter18,Edilson21}. 
From the SDSS Main Galaxy Sample, we use colour-colour diagrams to select the blue galaxies sample, with redshifts $z \in [0,0.30]$ and $z_{\mbox{\footnotesize eff}} = 0.166$. The methodology adopted to calculate the 3D comoving distances follows the literature.


%
%

This work is organized as follows. In section \ref{sec:Data} we present our data selection from our blue galaxies sample, i.e., the redshift range and the sky region to perform the analysis. Still in section \ref{sec:Data}, we explain the pipeline to construct the log-normal simulations and random catalogues, important ingredients to perform the 2PCF. In section \ref{sec:methodology} we describe our statistical tools to obtain the 2PCF. In sections \ref{sec:results} and \ref{sec:conclusions} we show our main results and the final remarks, respectively.

\section{Data sample and Simulations}\label{sec:Data}

In this section, we provide a comprehensive overview of the fundamental aspects pertaining the SDSS blue galaxies sample, the simulations employed for constructing the covariance matrix, and the random catalogues utilized for calculating the 2PCF.

\subsection{SDSS blue galaxies sample}\label{subsec:blue_sample}

We used the blue star-forming galaxies catalogue analyzed in~\citep{Avila19,Edilson21,Dias23}. It was selected blue star-forming galaxies from the galaxy colour-colour diagram, using the u, g, and r Sloan Digital Sky Survey (SDSS) broad bands~\cite {York00}. The SDSS magnitudes for each galaxy is corrected by Galactic extinction following~\cite{Schlegel98}. Also a k-correction was applied~\citep{Chilingarian10,Chilingarian12}. For further details on the data selection, see \cite{Avila19}.

\begin{figure}
\centering
\includegraphics[scale=0.45]{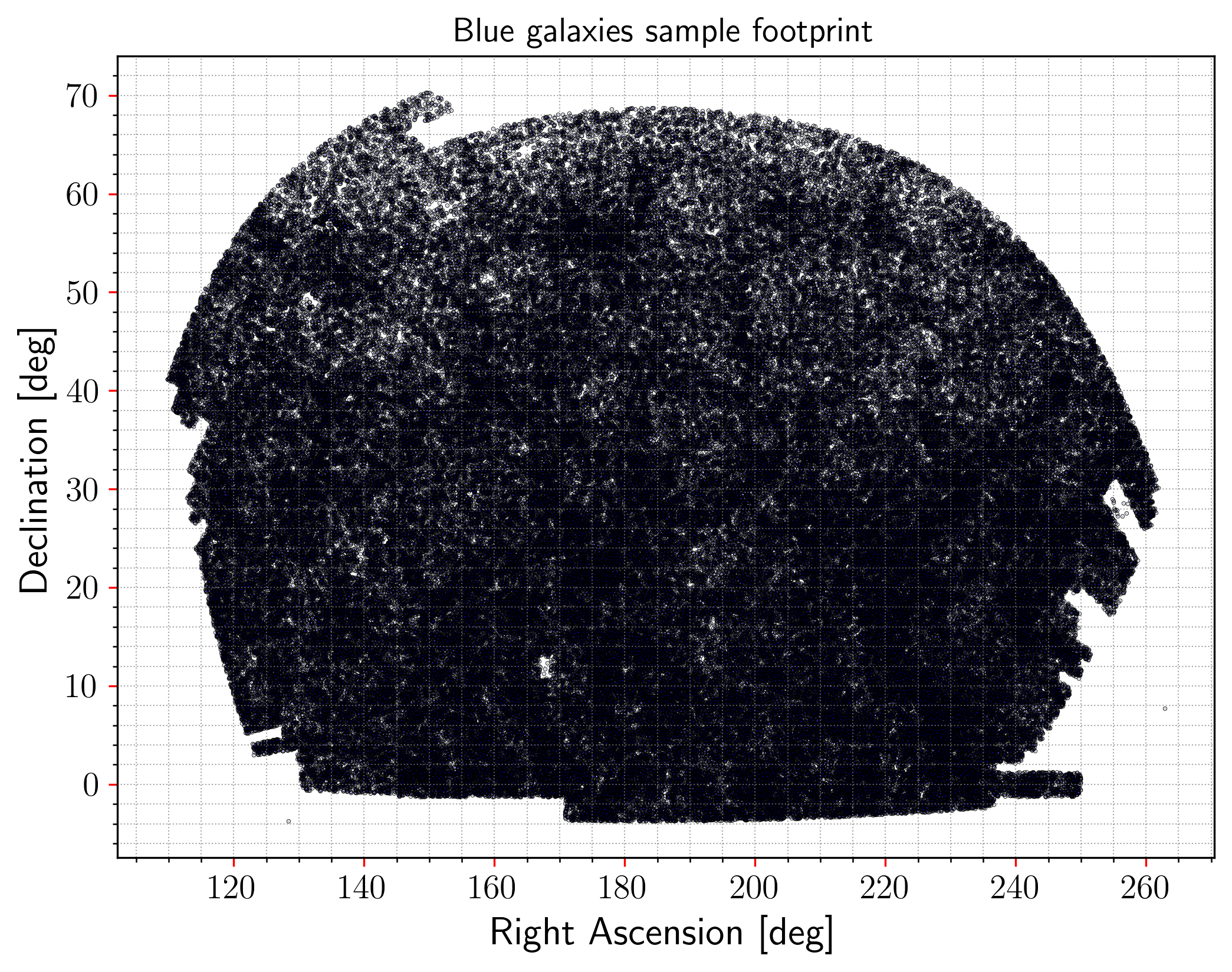}
\caption{The sky coverage of our sample of SDSS blue galaxies, located in the North Galactic hemisphere.}
\label{fig:sky_footprint}
\end{figure}

\begin{figure}
\centering
\includegraphics[scale=0.45]{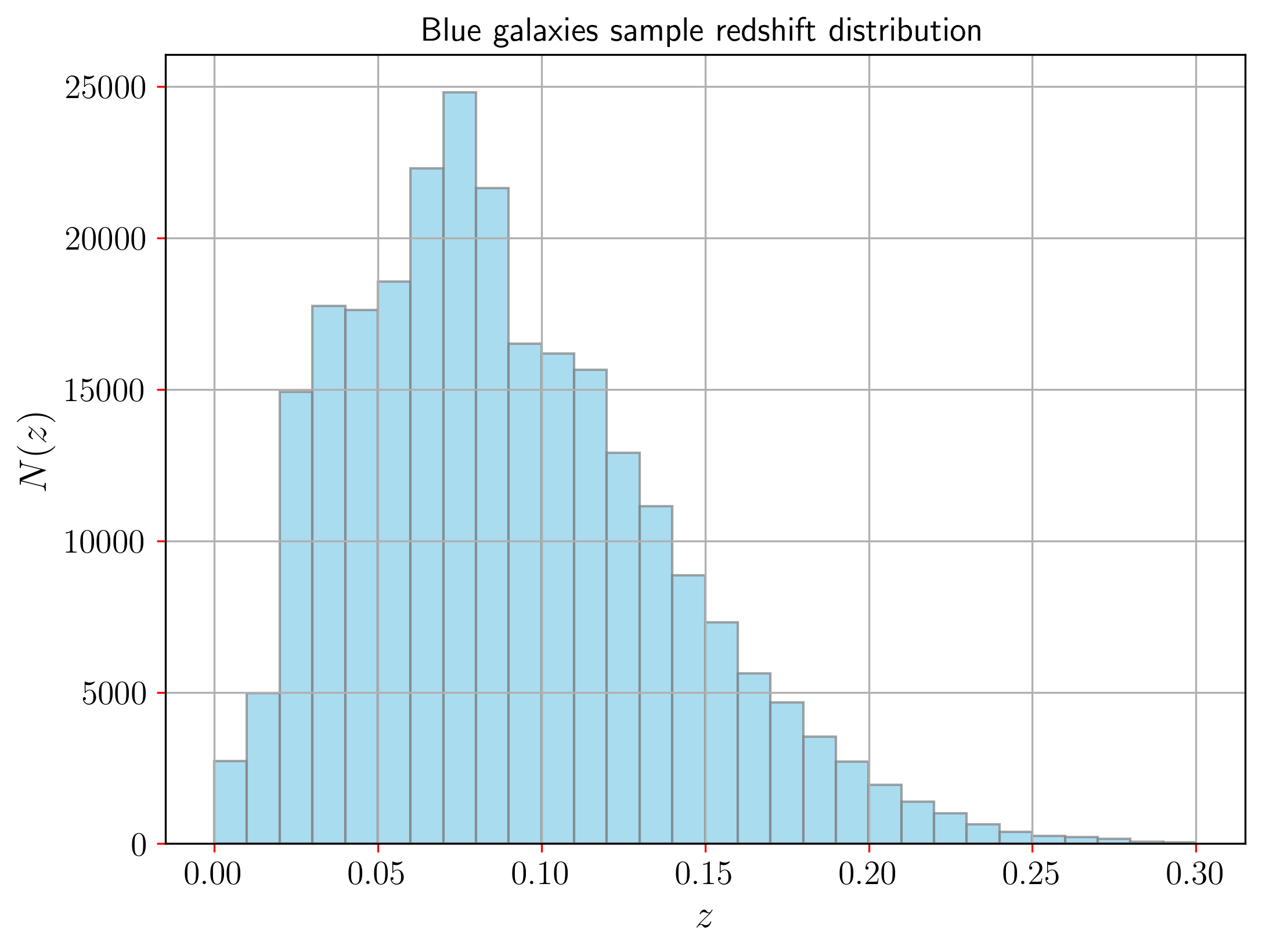}
\caption{The redshift distribution of the  
SDSS blue galaxies sample selected for our BAO analyses.}
\label{fig:redshift_distribution}
\end{figure}

To achieve statistical significance and successfully detect the BAO signature, a large effective volume is required, along with a high number density~\citep{Eisenstein05}. Despite the high numerical density of our sample, $\Bar{n}=6.4\times 10^{-3} ~h^{3}/{\text{Mpc}}^{3}$, the total sample volume remains limited. To enhance the statistical significance of the Baryon Acoustic Oscillations (BAO) signal, we decide to utilize almost the entire available sample. 
Specifically, we have selected the blue galaxies in the North Galactic Sample within the redshift range of $0 < z < 0.30$, totaling 256,478 objects. 
In figure~\ref{fig:sky_footprint} we show the sky footprint, 
covering an area of $\sim$ 7,000 deg$^2$, 
and in figure~\ref{fig:redshift_distribution} we observe the redshift distribution of our selected sample for analysis, respectively. 


Although we have included the entire sample, it is uncertain beforehand whether the BAO signal will be detected. One preliminary approach to determine if our sample size is optimal for BAO signal detection is through the calculation of the effective volume, defined as~\citep{Tegmark97}:
\begin{equation}\label{equation:effective_volume}
    V_{\text{eff}} \,\equiv\, 4\pi f_{\text{\footnotesize sky}} \int dz \left[\frac{n(z)P_{0}}{1+n(z)P_{0}}\right]^{2},
\end{equation}
where $f_{\text{\footnotesize sky}}$ is the sky fraction observed, which is  $1/6$ for our blue galaxies sample, and $P_{0}$ is the characteristic power spectrum amplitude of the BAO signal. 
We adopt $P_{0} = 10,000 \,\text{Mpc}^{3}/h^{3}$. 
For our blue galaxies sample, we have determined the effective volume to be $V_{\text{eff}} = 0.35 \,\text{Gpc}^{3}/h^{3}$. 
This value can be compared to significant studies such as the breakthrough work by \cite{Eisenstein05}, where they detected the BAO signal with an effective volume of $V_{\text{eff}}=0.38 \,\text{Gpc}^{3}/h^{3}$. 
Moreover, in comparison to BAO measurements in the Local Universe, such as~\cite{Beutler11} and~\cite{Carter18}, our effective volume proves to be highly capable of detecting the BAO signal. 

To conclude regarding the data selection, we incorporated weights to our blue galaxies sample using the FKP (Feldman-Kaiser-Peacock) weight procedure~\citep{Feldman94} 
\begin{equation}\label{equation:FKP_weight}
    w(z) = \frac{1}{1+n(z)P_{0}},
\end{equation}
where $n(z)$ represents the number density at redshift $z$. 
The incomplete sky survey, mainly due to observational characteristics, 
restricts the large-angular scrutiny of the universe. 
But this weighting procedure aims to minimize the variance in the 2PCF. 
Then, the effective redshift, $z_{\text{eff}}$, of our data sample in analysis can be calculated 
\begin{equation}\label{equation:effective_redshift}
    z_{\text{eff}} = \frac{\sum_{i} w_{i}z_{i}}{\sum_{i} w_{i}} \,,
\end{equation}
finding $z_{\text{eff}} = 0.166$.

\subsection{Log-normal simulations}\label{subsec:log_simulations}

The statistical significance of the BAO signal relies on calculating the covariance between measurements of the 2PCF on different scales~\citep{Eisenstein05,Cole05,Beutler11,Carter18}. 
It is crucial to assess the uncertainties and correlations inherent in these measurements. 
In addition, it plays a vital role in distinguishing genuine BAO signals from random fluctuations. 
Moreover, the covariance matrix allows to evaluate the impact of various observational effects on BAO measurements, such as sample selection and survey geometry. 

Generally, the determination of the covariance matrix involves constructing simulated catalogues that mimic the characteristics of the actual catalogue under study~\citep{Springel05,Vogelsberger14}. In this work, we employ log-normal simulations~\citep{Coles91}, which model the density field as a log-normal random field. This approach allows us to replicate the statistical properties of the observed blue galaxies distribution and generate synthetic catalogues that closely resemble the real data~\citep{Marulli16,Xavier16,Agrawal17,Hand18,Ramirez22}. 
Recent works indicate that, in terms of the primary statistical estimators for galaxy distribution analysis (correlation function, power spectrum, and bispectrum), both N-body simulations and log-normal simulations yield comparable results for the covariance matrix~\citep{Lippich19, Blot19, Colavincenzo19}.

In this work, we build 1000 simulated log-normal catalogues with the public code presented in \cite{Agrawal17}\footnote{\url{https://bitbucket.org/komatsu5147/lognormal_galaxies}}. 
Recently, we successfully implemented this code in our study to examine the gravitational dipole in the Local Universe~\citep{Avila21}. 
The input parameters needed to generate the mock catalogues that reproduce our blue sample clustering features are listed in Table~\ref{tab:table1}. They are the redshift $z$ (median redshift of the catalogue), the bias $b$, the number of galaxies $N_g$ and the box dimensions\footnote{In units of Mpc/$h$.} ($L_x, L_y, L_z$). We set a $313\times512\times311$ grid for the Fourier transformation given a total resolution of order $\sim 2.4$ Mpc/$h$. The nonlinear matter power spectrum (Halofit) used as an input in the code is obtained with the \textsc{camb}\footnote{\url{https://lambda.gsfc.nasa.gov/toolbox/camb_online.html}} tool~\citep{CAMB}, calculated at $z=0.08$. Finally, to obtain the correlation function in the redshift space, the positions of galaxies are shifted by the velocity in the $z$ - direction\footnote{The choice of coordinate does not affect the final measurement since, in the analysis in question, we are only looking at the projection of the velocity field.}.

\begin{table}
\caption{Survey configuration and cosmological parameters from the Planck last data 
release~\citep{Planck20} used to generate the set of $N_{\mbox{s}} = 1000$ log-normal realisations used in the analyses.}
\centering
\setlength{\extrarowheight}{0.2cm}
	\begin{tabular}{c|c}
	\hline
	Survey configuration          & Cosmological parameters         \\ 
\hline
	$z=0.08$                              & $\Omega_{c}h^{2}=0.1202$       \\  
	$b=1.1$                              & $\Sigma m_{\nu}=0.0600$         \\
	$N_{g}=4.6\times 10^{5}$     & $n_{s}=0.9649$                         \\
	$L_{x}=746$                     & $\ln(10 A_{s})=3.045$                  \\
	$L_{y}=1222$                     & $\Omega_{b}h^{2}=0.02236$       \\
	$L_{z}=741$                     & $h=0.6727$       \\ \hline                          
	\end{tabular}
\label{tab:table1}
\end{table}

The code generates a simulated catalogue in Cartesian coordinates with predetermined dimensions. However, to ensure that these simulations accurately represent the observational conditions, we apply specific cuts to both the geometry and the distribution of points along the distance. These cuts ensure that the final 1000 catalogues possess the same footprint and redshift distribution as the blue galaxy sample. Finally, we apply the weights to each simulated galaxy.

\subsection{Random catalogues}\label{subsec:random_catalogues}

Random catalogues, also known as uncorrelated catalogues with similar characteristics to the catalogue under study, are of significant importance in obtaining a reliable measurement of the 2PCF~\citep{Keihanen19}.

In this study, we employ two equivalent pipelines for constructing random galaxy catalogues. Specifically, we constructed the random catalogue in spherical coordinates for the data and in Cartesian coordinates for the simulations. This methodology was adopted to eliminate the need for coordinate transformations, thereby ensuring the integrity and precision of our results. 

For the blue galaxies sample, we utilize the publicly available \textsc{randomsdss}\footnote{\url{https://github.com/mchalela/RandomSDSS}}, which provides access to extensive SDSS maps, enabling us to capture the survey geometry accurately. This code also generates a random redshift distribution based on a given sample.

Regarding the log-normal simulations, we employ the numpy code~\citep{Harris20} to generate a uniform distribution within a 3D rectangular shape defined by the dimensions specified in 
Table~\ref{tab:table1}. Subsequently, using the same code applied to the simulations, we extract a random catalogue that possesses the same characteristics as the blue galaxy sample. 

Notably, both random catalogues consist of five times more data points than the blue galaxies sample~($N \simeq 1.3 \times 10^{6})$, ensuring minimal noise in the correlation function. Also, similar to the approach employed for the blue galaxies sample and simulations, we assigned weights to each object within the random catalogue using the FKP weight procedure, as defined by equation (\ref{equation:FKP_weight}).

\section{Methodology}
\label{sec:methodology}

The BAO methodology entails nuanced considerations, particularly regarding the estimation and modeling of the two-point correlation function (2PCF). Two key elements warrant special attention: the choice of the correlation function estimator~\citep{Vargas13} and the level of sophistication in the theoretical model used to describe the correlation function across various scales of analysis~\citep{Crocce08}. This includes considerations for non-linear, quasi-linear, and linear regimes, each contributing to the comprehensive characterisation of BAO.
In the following section, we will delineate the methodology employed in this study. This encompasses the selection of the 2PCF estimator, the cosmological framework utilized for distance calculations, the determination of the covariance matrix, and the empirical models employed for inferring cosmological parameters. It is crucial to emphasize that, in modeling the 2PCF, we have deliberately avoided adopting a specific physical model in this work.

\subsection{The 2-point correlation function}\label{subsec:2PCF}

The investigation of galaxy clustering, as well as other cosmological tracers, commonly relies on the analysis of the 2PCF~\citep{Peebles74,Landy93}. This statistical tool serves as a fundamental approach for studying the spatial distribution and clustering properties of galaxies in the universe. For alternative approaches to clustering analysis, see, 
e.g.,~\cite{Sanchez11,Carnero12,Avila18,Marques18,Marques20,Edilson20}.

The 2PCF is obtained by counting pairs of cosmic objects in a data set at a given separation 3D distance $s$, $DD(s)$, and pairs of simulated objects from a random set, $RR(s)$. The most used 2PCF estimator in astrophysical applications is the Landy-Szalay \citep[LS;][]{Landy93}, because it returns the smallest deviations for a given cumulative probability, besides having no bias and minimal variance~\citep{Kerscher00}. This estimator is defined by 
\begin{eqnarray}\label{equation:LS_estimator}
\xi(s) \equiv \frac{DD(s) - 2DR(s) + RR(s)}{RR(s)} \, ,
\end{eqnarray} 
where $DR(s)$ counts the pairs, with one object in the data set and the other in the random set, separated by a distance $s$. The quantity $\xi(s)$ gives the excess probability of finding two points of a data set at a given separation distance $s$ when compared to a random distribution. The measurements of $\xi(s)$ were obtained using the code \textsc{treecorr}~\citep{Jarvis04}\footnote{\url{https://github.com/rmjarvis/TreeCorr}}.

The distances $s$ between galaxies $i$ and $j$ making an angle of $\theta_{ij}$ can be obtained using the following expression
\begin{equation}\label{separation_distance}
    s(z_{i}, z_{j}, \theta) = \sqrt{\chi(z_{i})^{2} + \chi(z_{i})^{2} - 2\chi(z_{i})\chi(z_{j})\cos{\theta_{ij}}},
\end{equation}
where $\chi(z)$ is the comoving distance for a galaxy with redshift $z$. In general, a standard approach is to adopt a fiducial cosmological model, like $\Lambda$CDM model, and it is fixed, i.e., we do not repeat the correlation function for other models or parameters. Such a procedure can bias the shift parameters if we do not use the correct model~\citep{Carter20,Anselmi22,He23}.

As we are using a sample of galaxies at very low redshift, it becomes interesting to test different models of cosmology to convert $z$ to $s$. We chose to work with four cosmological models, hereafter termed {\em samples}: 
\begin{itemize}
    \item Sample 1: The flat-$\Lambda$CDM model~($\Omega_{k,0}=0$) with 
          \begin{equation}\label{eq:sample_I}
              \chi(z) \equiv \frac{c}{H_{0}}\int_{0}^{z}\frac{dz'}{E(z')},
          \end{equation}
          where $E(z)=\sqrt{\Omega_{m,0}(1+z)^{3} + 1 - \Omega_{m,0}}$. We adopt Planck-$\Lambda$CDM values in the parameters $\Omega_{m,0}$ and $H_{0}$~(see table \ref{tab:table1}, second column). 
          
    \item Sample 2: Cosmography in first order expansion~\citep{Visser05} using the Hubble-Lemaître relation,
          \begin{equation}\label{eq:sample_II}
              \chi(z) = \frac{cz}{H_{0}}.
          \end{equation}
          
    \item Sample 3: Cosmography in third order expansion,
          \begin{equation}\label{eq:sample_III}
              \chi(z) = \frac{cz}{H_{0}}\left[1 - \left(1+\frac{q_{0}}{2}\right)z + \left(1+q_{0}+\frac{q_{0}^{2}}{2}-\frac{j_{0}}{6}\right)z^{2}\right],
          \end{equation}
          where we set the parameters ($q_{0}, j_{0}$) to (-0.55, 1.00)~\citep{Li20}. That is, parameters fixed to mimic a Planck-$\Lambda$CDM cosmographic expansion.
          
    \item Sample 4: Cosmography in third order expansion for the set ($q_{0}, j_{0}$) in (-0.71, 1.26)~\citep{Liu23}. Thus, a small deviations in cosmographic distances from Planck-$\Lambda$CDM values, but compatible with current error bars on these parameters.
\end{itemize}

\begin{table}
\centering
\caption{Summary information of the best-fit at 1$\sigma$ CL for the BAO scale, $s_{_{\text{BAO}}}$, for the four samples.}
\label{tab:bao_fit}
\setlength{\extrarowheight}{0.2cm}
\begin{tabular}{ccc}
    \hline
	Methodology  & $s_{_{\text{BAO}}}[\text{Mpc}/h]$ & $s_{_{\text{BAO}}}[\text{Mpc}/h]$ (with $\alpha$ free)\\ 
	\hline
    Sample 1 & $100.28 ^{+10.79} _{-22.96}$ & $ 88.92^{+38.06}_{-32.59}$ \\ 
    Sample 2 & $108.94 ^{+13.69} _{-16.70}$ & $ 92.76^{+35.49}_{-30.78}$ \\
    Sample 3 & $87.28 ^{+9.26} _{-21.86}$ & $ 80.87^{+35.64}_{-29.64}$  \\ 
	Sample 4 & $89.21 ^{+9.24} _{-23.02}$ & $ 84.49^{+35.06}_{-30.60}$ \\ 
	\hline
    \end{tabular}
\end{table}

\begin{figure}
    \centering
    \includegraphics[scale=0.6]{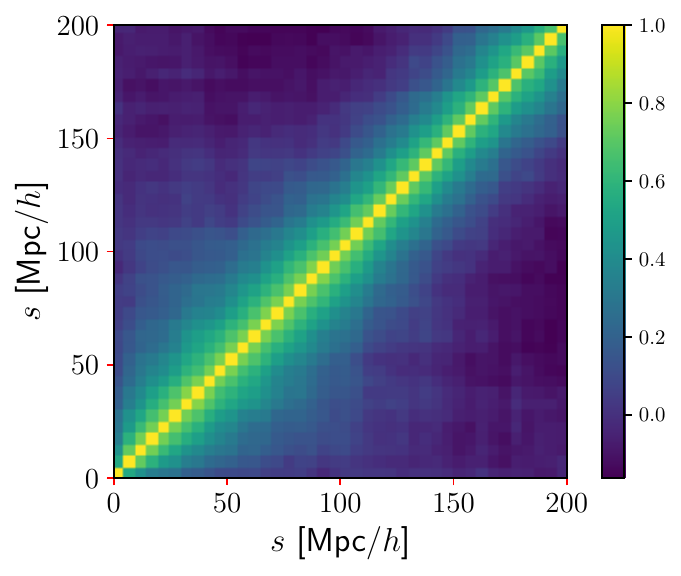} 
    \caption{Correlation matrix derived from a covariance matrix calculated from 1000 log-normal realisations (see section~\ref{subsec:random_catalogues} for details).}
    \label{fig:cov}
\end{figure}




\begin{figure*}
    \centering
    \includegraphics[scale=0.5]{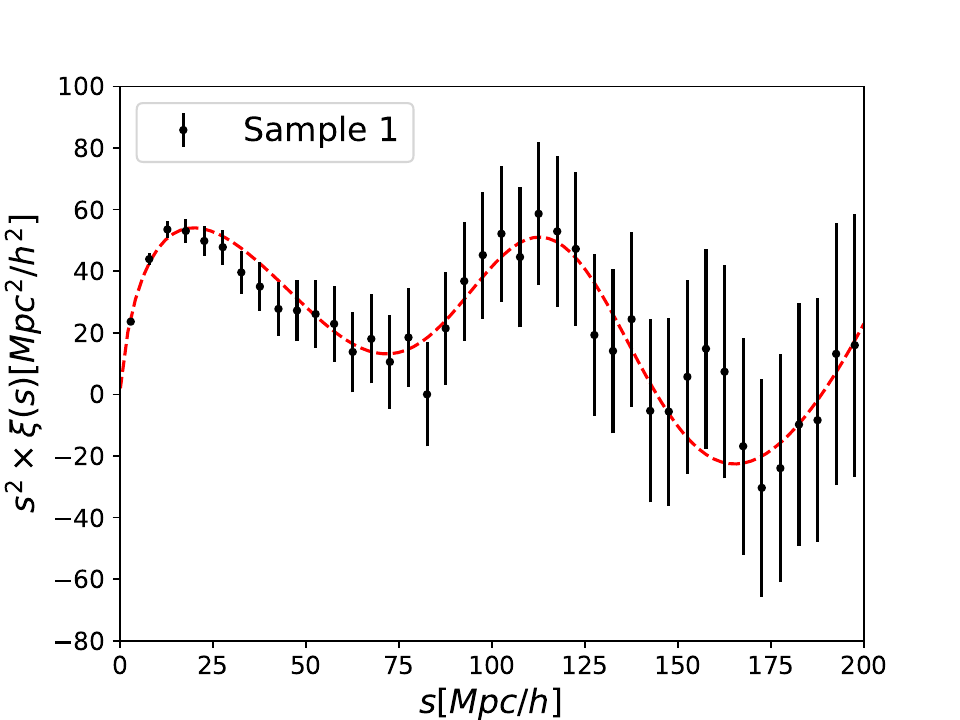} \,\,\,\,\,\,
    \includegraphics[scale=0.5]{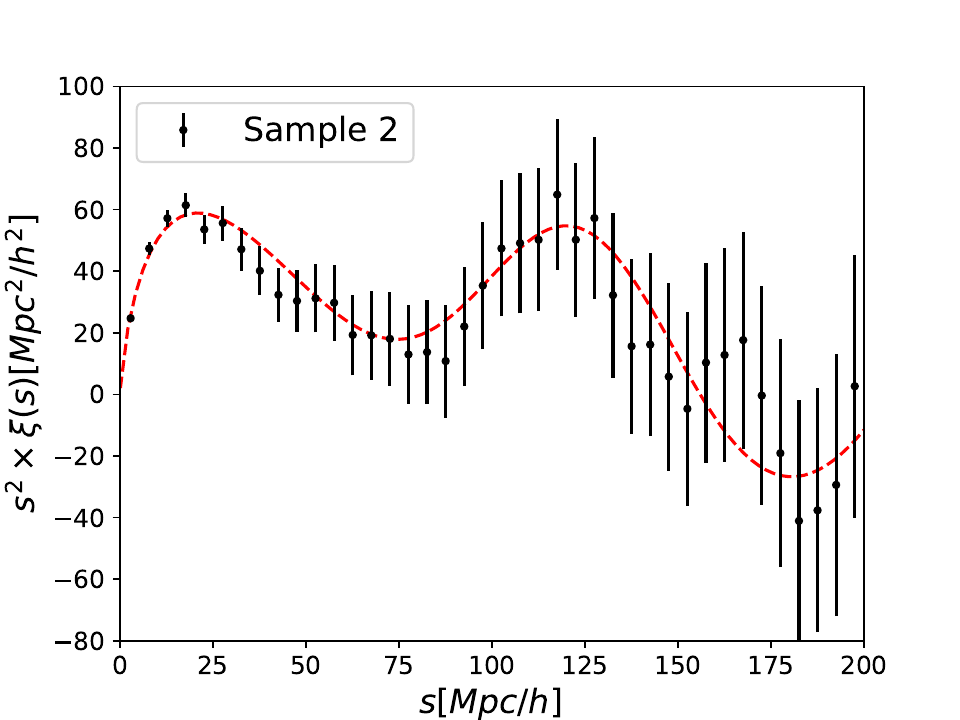}
    \caption{Left panel: BAO signature obtained in the 2PCF by analyzing the SDSS blue galaxies sample in the redshift interval z $\in$ [0, 0.30] within the perspectives of sample 1. Right panel: Same as in the left panel, but for the sample 2.}
    \label{fig:best_fit_models_1_2}
\end{figure*}

\begin{figure*}
    \centering
    \includegraphics[scale=0.5]{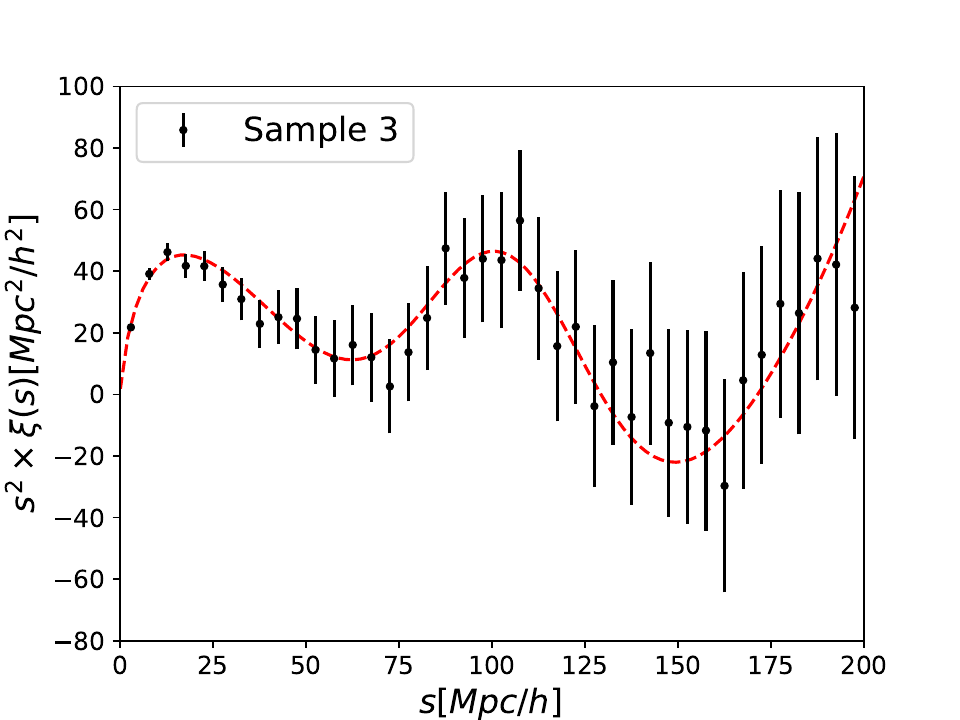} \,\,\,\,\,\,
    \includegraphics[scale=0.5]{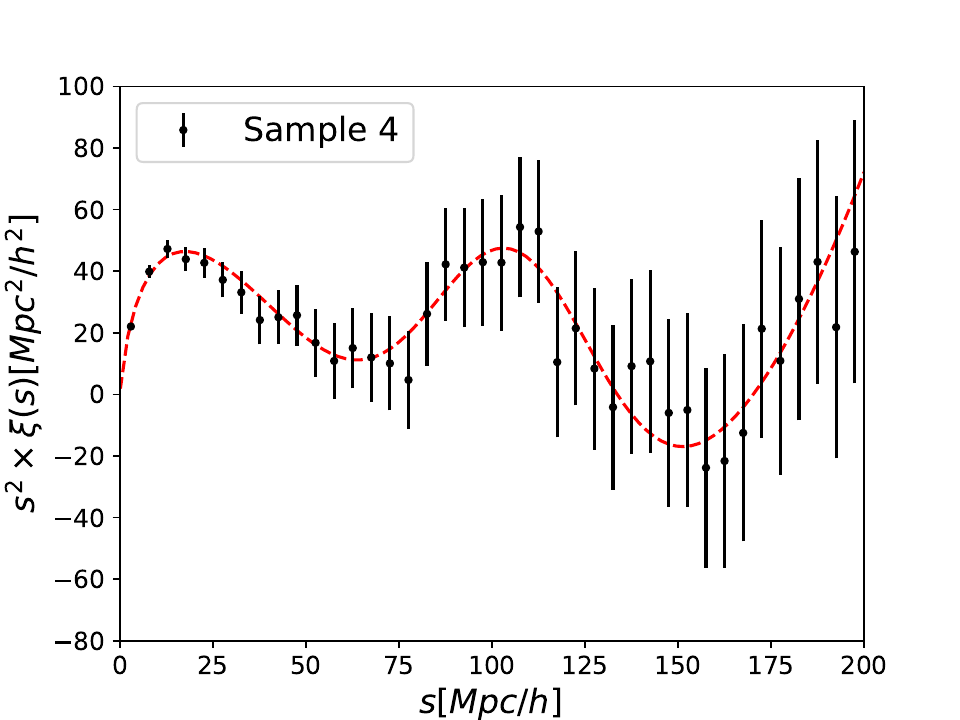}
    \caption{Left panel: BAO signature obtained in the 2PCF by analysing the SDSS blue galaxies sample in the redshift interval z $\in$ [0, 0.30] within the perspectives of sample 3. Right panel: Same as in the left panel, but for the sample 4.}
    \label{fig:best_fit_models_3_4}
\end{figure*}

\begin{figure}
\centering
\includegraphics[scale=0.5]{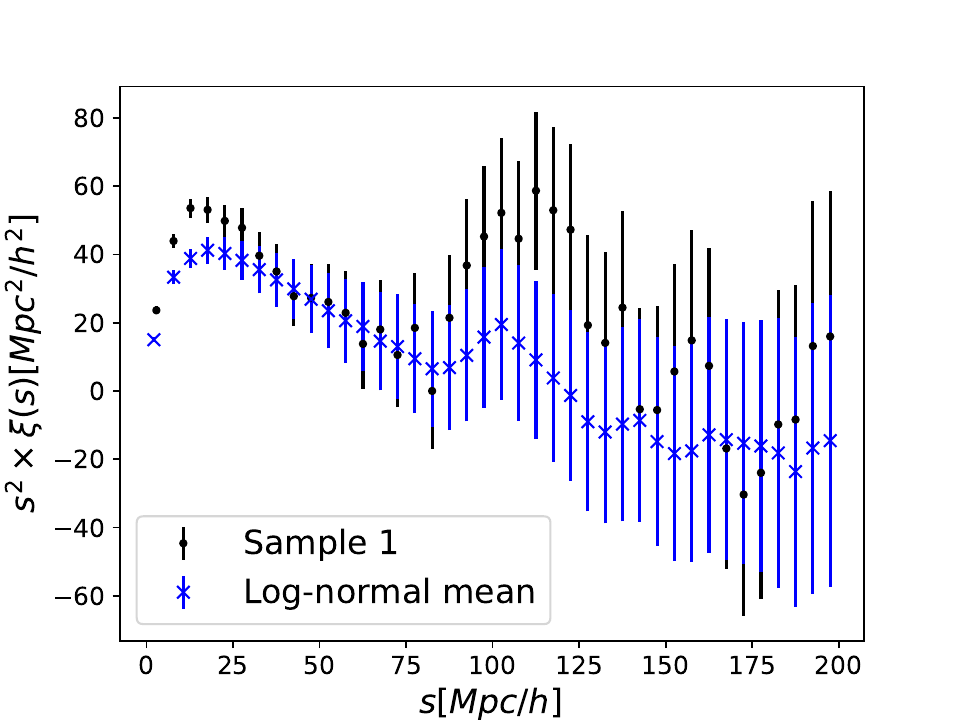} 
\caption{The 2PCF, represented by 
multiplication signs \textsf{x}, corresponds 
to the mean value from the $1000$ log-normal 
simulations used to calculate the covariance 
matrix. 
For comparison, the black dots correspond to the analysis of the correlation function for Sample 1 
(see the text for more details).}
    \label{fig:mock_data}
\end{figure}

\begin{figure}
    \centering
    \includegraphics[scale=0.5]{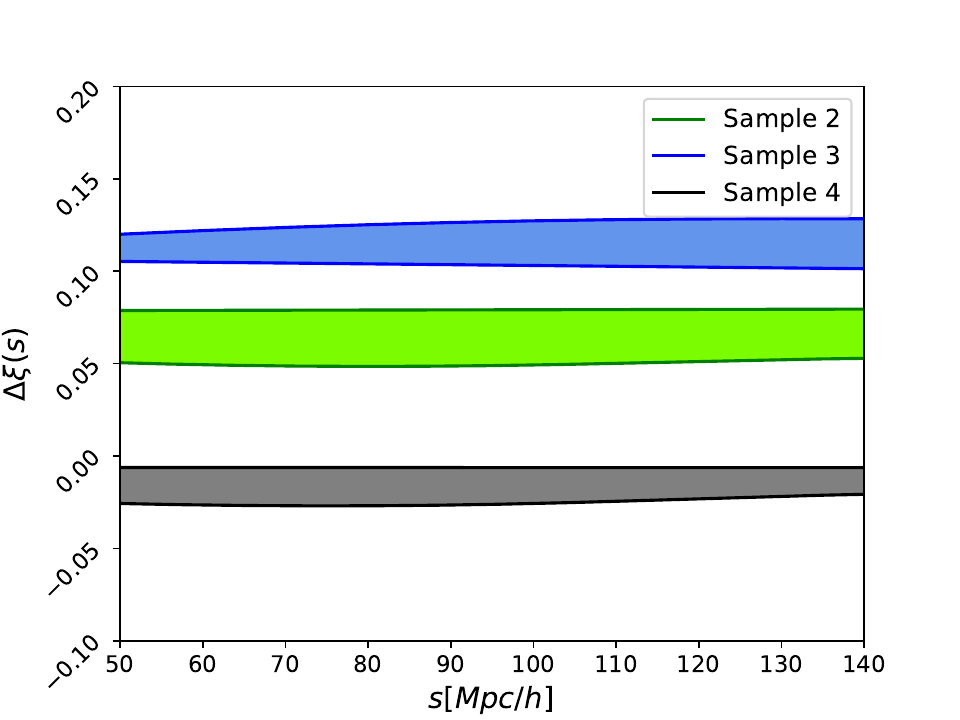} 
    \caption{Relative difference in the 2PCF fit at 1$\sigma$ CL for the samples 2, 3, and 4, with respect to the predicted by the sample 1.}
    \label{fig:delta_s}
\end{figure}

To estimate the covariance matrix and the significance of our results, we have used a set of $N=1000$ galaxy mocks described above (see the subsection~\ref{subsec:log_simulations}). For each mock, we extract the information about the 2PCF for the 3D case. The covariance matrix for $\xi(s)$ was estimated using the expression
\begin{eqnarray}\label{equation:cov_matrix}
\mbox{Cov}_{ij} = 
\frac{1}{N}\sum_{k=1}^{N}\left[ \mbox{X}_k(i)-\overline{\mbox{X}}(i)\right]
                \left[\mbox{X}_k(j)-\overline{\mbox{X}}(j)\right] \, ,
\end{eqnarray} 
where the $\mbox{X}_k(i)$ term represents the statistics used, that is, $\xi$ for the 2PCF, in the $i$-th bin, $i=1,\dots,N_b$ for the $k$-th mock, $k=1,\dots,N$; $\overline{\mbox{X}}(i)$ is the mean value for this statistics over the $N = 1000$ mock samples in that bin. Finally, the error of $\mbox{X}(i)$ is the square root of the main diagonal, 
$\delta \mbox{X}(i)= \sqrt{\mbox{Cov}_{ii}}$.

To fit the data from our correlation function 
analyses, we employ an empirical model to 
characterise the 2PCF in redshift space, as 
described in previous works~\citep{Sanchez11, Carnero12, Edilson18, Edilson21}. 
Additionally, we account for the dilation of 
distances, a consequence of potentially inaccurate 
fiducial cosmology choices~\citep{Heinesen_2019} 
\begin{eqnarray}
\label{eqFIT3D}
\xi(s) = A + B \, s^{\,\delta} + C \exp^{-(s - s_{BAO})^2 / 
2 \Sigma^{2}} \,+\, D \,s^{-1} .
\end{eqnarray}
In the given parameterization: $A$, $B$, $\delta$, $C \rightarrow \alpha^2 C$, $D$, $s_{\text{BAO}} \rightarrow \alpha^2 s_{\text{BAO}}$, and $\Sigma \rightarrow \Sigma/\alpha$ are free parameters, where the isotropic scaling parameter, denoted by $\alpha$, plays a crucial role. 
To ensure an accurate representation of the scales close to the BAO signature, we adopted a Gaussian function (the third term in equation~(\ref{eqFIT3D})) to model it.




We use the Markov Chain Monte Carlo (MCMC) method to analyse the parameters $\theta_i = \{A, B, \delta, C, D, s_{\text{BAO}}, \Sigma \}$, building  the posterior probability distribution function 

\begin{equation}
\label{L}
 p(D|\theta) \propto \exp \Big( - \frac{1}{2} \chi^2\Big) \, ,
\end{equation}
with
\begin{equation}
\chi^2 = [\xi_{\rm th}(\theta) - \xi_{\rm obs}]^T\mbox{Cov}^{-1}[\xi_{\rm th}(\theta)  - \xi_{\rm obs}],
\end{equation}
where we note the 2PCF data, $\xi_{\rm obs}$, and the model (theory), $\xi_{\rm th}$, as a function of parameters $\theta$; \,$\mbox{Cov}^{-1}$ means the inverse of the covariance matrix.

The objective of any Markov Chain Monte Carlo (MCMC) approach is to generate $N$ samples $\theta_i$ from the general posterior probability density given by:
\begin{equation}
\label{psd}
p(\theta_i, \alpha|D) = \frac{1}{Z} p(\theta,\alpha) p(D|\theta,\alpha).
\end{equation}

Here, $p(\theta,\alpha)$ and $p(D|\theta,\alpha)$ represent the prior distribution and the likelihood function, respectively. The variables $D$ and $\alpha$ correspond to the set of observations and potential nuisance parameters. The term $Z$ denotes a normalisation factor. Our statistical analysis employs the \textit{emcee} algorithm \citep{Foreman-Mackey_2013}. The priors on the baseline parameters are assumed as follows: $A \in [-5, 5]$, $B \in [0, 600]$, $\delta \in [-5, 5]$, $C \in [-1, 1]$, $D \in [-700, 0]$, $s_{\text{BAO}} \in [30, 150]$, $\Sigma \in [10, 100]$, and $\alpha \in [0.1, 2.0]$.

\section{Main Results}
\label{sec:results}

In the following, we systematically explore the entire parameter space to constrain the probability distribution of key parameters, with a primary focus on fitting the Baryon Acoustic Oscillation (BAO) scale within our datasets. Table \ref{tab:bao_fit} provides a summary of the best-fit values at the 1$\sigma$ confidence level (CL) for the BAO scale, $s_{\scalebox{0.65}{\rm BAO}}$, across the four samples. Our investigation begins with sample 1, the Planck-$\Lambda$CDM sample. 
Considering $\alpha = 1$ we obtain $s_{\scalebox{0.6}{\rm BAO}} = 100.28^{+10.79}_{-22.96}$ Mpc$/h$, but considering $\alpha$ as a free parameter yields 
$s_{\scalebox{0.6}{\rm BAO}} = 88.92^{+38.06}_{-32.59}$, which indicate a precision of 16\% and 38\%, respectively 
(see Table~\ref{tab:bao_fit_ps} for a summary of all constraints). 
In Appendix \ref{appendix} we present the 1- and 2-dimensional projections of the posterior probability distributions across the entire parametric space for the four samples examined in this study. Additionally, we summarize the observational constraints in Tables \ref{tab:bao_fit_ps} and \ref{tab:bao_fit_ps_alfa_free}.
Expanding our approach to fitting and investigating the BAO scale using alternative distance measurements (with $\alpha = 1$) in samples 2, 3, and 4, we obtain 
a precision of 18\%, 14\%, and 18\%, respectively. However, treating $\alpha$ as a free parameter leads to increased degeneracy in the space of fitting parameters from a statistical standpoint. Furthermore, a notable correlation between $\alpha$ and $s{\scalebox{0.6}{\rm BAO}}$ is observed. 
Consequently, analyses with $\alpha$ as a free parameter result 
in a natural decrease in our estimate precision. 
Although it is observed slight deviations in average values, all measurements exhibit statistical equivalence to each other at less than the 1$\sigma$ confidence level.

Figure \ref{fig:cov} presents the correlation matrix derived from the covariance matrix for the two-point correlation function (2PCF), utilizing the set of log-normal simulated maps developed in this study. In Figures \ref{fig:best_fit_models_1_2} and \ref{fig:best_fit_models_3_4}, we illustrate the theoretical curve fitting by the BAO empirical model for the four 2PCF samples under consideration. 
To quantify the differences, we calculate the quantity
\begin{equation}
{\Delta \xi}_i \equiv \frac{\xi(s)_{\rm sample \, 1}-\xi(s)_{\rm sample \,  i}}{\xi(s)_{\rm sample \, 1}} \,,
\end{equation}
where $i$ varies across samples 2, 3, and 4. This measure directly indicates the relative difference in the 2PCF of each sample concerning the best-fit values obtained from samples 2, 3, and 4, concerning the results provided by sample 1.

Following \cite{Beutler11}, we compare the log-normal simulations and the data sample. 
In Figure \ref{fig:mock_data} we show, as blue multiplication signs \textsf{x}, the mean of 1000 log-normal simulations. The black dots represent the data points for the Sample 1. As expected from the log-normal simulations, we found good agreement at intermediate scales, $30 < r < 80$ Mpc$/h$. 
On small scales, i.e., $r \lesssim 30$ Mpc$/h$, and close to the BAO scale, the log-normal simulations do not correctly capture the $\xi(s)$ amplitude. 
Such limitation in this set of simulations is already expected~\citep{Agrawal17}. 
These limitations are quantified when computing the errors in the correlation function parameters.

Figure \ref{fig:delta_s} illustrates ${\Delta \xi}i$ as a function of distances $s$ within the 1$\sigma$ reconstruction range. Notably, over the scales of interest ranging from 
$50$ Mpc$/h$ to 140 Mpc$/h$, the samples exhibit an average difference of approximately $\sim$ 4\%, $\sim$ 9\%, and $\sim$ 4\% for samples 2, 3, and 4, respectively, relative to sample 1. For additional details, including the comprehensive exploration of the parametric space for our baseline parameters and tables containing the best-fit values for all free parameters in our analysis, please refer to Appendix~\ref{appendix}.

\subsection{Cosmological Interpretation}

Within the homogeneous and isotropic universe (see, e.g.,~\citet{Dias23,Kester23,Franco24}), and under the $\Lambda$CDM framework assumption, the BAO bump in redshift space can be evaluated by~\citep{Bassett10} 
\begin{equation}\label{Deltaz}
\Delta z_{\scalebox{0.6}{\rm BAO}}(z) 
= \frac{s_{\scalebox{0.65}{\rm BAO}} \,H(z)}{c} \,,
\end{equation}
where $H(z)$ is the Hubble function and $c$ is the speed of light. 
Using the fiducial cosmology of our sample 1, as well as our inferred value for $s_{\scalebox{0.65}{\rm BAO}}$, we find 
\begin{equation}
\Delta z_{\scalebox{0.6}{\rm BAO}}(z_{\rm eff}=0.166) \,=\, 0.0361^{+0.00262}_{-0.0055} \,,
\end{equation}
at 1$\sigma$~CL.

We emphasise that our estimate above for the 
$\Delta z_{\scalebox{0.6}{\rm BAO}}$ is a model-dependent geometrical quantity because it is necessary to assume an input cosmology to infer $H(z)$, but the BAO scale $s_{\scalebox{0.65}{\rm BAO}}$ can be obtained in a model-independent way, as we did in our sample 2. 
The discriminant $\Delta z_{\scalebox{0.6}{\rm BAO}}$ can be used to constraint dark energy models in the Local Universe, as this parameter is sensitive to dynamics in the $H(z)$ function~\citep{Benisty21,Staicova22,DAgostino23,Bikash23,Akarsu23,Benisty23,Giare23}.

Our result is a new independent measurement of BAO in the Local Universe, with a sample of SDSS blue galaxies at low redshift. 
This allows us to compare our measurement with other results in similar redshift intervals, like the analyses of~\cite{Beutler11} 
and~\cite{Carter18} at the effective redshifts $z_{\text{eff}}=0.106$ and 
$z_{\text{eff}}=0.097$, respectively, where both works constrained the BAO scale studying the 6dF Galaxy Survey~\citep{Jones04}. 
The main difference between these analyses is the application of the density field reconstruction done by~\cite{Carter18}. 
Estimating $\Delta z_{\scalebox{0.6}{\rm BAO}}$ for both works, we obtain 0.0342 and 0.0315, for \cite{Beutler11} and \cite{Carter18}, respectively. 
Complementing these data, \cite{Marra_2019} measured 
$\Delta z_{\scalebox{0.6}{\rm BAO}} = 0.0456 \pm 0.0042$ at $z = 0.51$
using a different methodology and data sample. 
In Table~\ref{tab:Delta_z} we display four measurements of 
$\Delta z_{\scalebox{0.6}{\rm BAO}}$ at diverse $z_{\text{eff}}$, 
where the radial BAO signature increases with the redshift as the universe expands~\citep{Bassett10}.

On the other hand, the parameter $\alpha$ allows us to fit the 2PCF obtained using a fiducial cosmological model (to calculate distances from redshifts) 
without the necessity to recalculate the 2PCF for every new set of cosmological parameters. 
The isotropic dilation parameter is defined as \citep{Eisenstein05,Beutler_2016}
\begin{equation}
\alpha \equiv \frac{D_V(z_{\text{eff}}) \,s^{\rm fid}_{\scalebox{0.6}{\rm BAO}} }{D_V^{\rm fid}(z_{\text{eff}}) \,s_{\scalebox{0.6}{\rm BAO}} } \,,
\end{equation}
where $D_V$ is the spherically-averaged distance. Taking into account our Sample 1 (where $\Lambda$CDM was used to calculate distances), we find $D_V(z_{\text{eff}}=0.166)/s_{\scalebox{0.6}{\rm BAO}} = 4.33^{+1.70}_{-1.40}$ at $1\sigma$ CL. Following standard procedure, we directly obtain the quantity $D_V(z_{\text{eff}}=0.166)/s_{\scalebox{0.6}{\rm BAO}}$ as a parameter derived from our chains. For $s^{\rm fid}_{\scalebox{0.6}{\rm BAO}}$, we assume the values predicted by Planck-$\Lambda$CDM cosmology~\citep{Planck20}. We note a significant decrease in precision compared to recent measurements conducted at high $z$ using other cosmic tracers \citep{Alam21}. 
This decrease can be attributed to the relatively smaller volume 
of our current sample in the Local Universe compared to other catalogues.

\begin{table}
\centering
\caption{
Measurements of $\Delta z_{\scalebox{0.6}{\rm BAO}} = 
\Delta z_{\scalebox{0.6}{\rm BAO}}(z_{\rm eff})$ from data surveys 
at different $z_{\rm eff}$, where the radial BAO signature increases with redshift.
}
\label{tab:Delta_z}
\setlength{\extrarowheight}{0.2cm}
\begin{tabular}{ccc}
\hline
$z_{\rm eff}$  & $\Delta z_{\scalebox{0.6}{\rm BAO}}$ & reference \\ 
\hline
0.097 & $0.0315$ & \cite{Carter18} \\ 
0.106 & $0.0342$ & \cite{Beutler11} \\
0.166 & $0.0361^{+0.00262}_{-0.00550}$ & this work \\
0.510  & $0.0456 \pm 0.0042$ & \,\,\cite{Marra_2019} \\
\hline
\end{tabular}
\end{table}

\section{Final Remarks}
\label{sec:conclusions}

BAO measurements have become one of the main cosmological tools nowadays. 
It is a fundamental probe for testing the physical nature of the dark components of the universe, i.e., dark energy and dark matter. 
In this scenario, new and independent BAO measurements made by diverse research teams, at different redshifts and precision, and most importantly, with diverse cosmological tracers, are needed for broader coverage of BAO measurements in the literature. 
In this work, we measure the BAO sound horizon scale 
$s_{\scalebox{0.65}{\rm BAO}} = 100.28^{+10.79}_{-22.96}$ Mpc$/h$, at the effective redshift $z_{\mbox{\footnotesize eff}} = 0.166$. 
This represents a measure of the BAO scale in the Local Universe with a precision of 16\%. 
Additionally, we performed another three BAO analyses according to different approaches that calculate radial distances from redshifts 
--all three are based on diverse cosmographic approaches--, as a way to check for possible biases or systematics in our methodology to measure the BAO scale, $s_{\scalebox{0.65}{\rm BAO}}$. 
As a result, we do not find significant deviations in our measurements, 
all in statistical agreement with this $s_{\scalebox{0.65}{\rm BAO}}$ measurement, as shown in Appendix \ref{appendix}. 

With the gravitational attraction acting during the longest period of cosmic time, $z \simeq 0$, the Local Universe is plenty of large overdense (superclusters) and large underdense (supervoids) regions~\citep{Courtois13,Hoffman17,Tully19,Avila22,Avila23}, structures that affect the computation of the BAO sound horizon scale~\citep{Crocce08}, $s_{\scalebox{0.65}{\rm BAO}}$, difficulty manifested in the large uncertainty obtained. 
Recently,~\cite{tully2023hooleilana} reported the discovery of a strong BAO signal at $z = 0.068$. 
In this sense, our results complement, and reinforce, the efforts to search for BAO signals in the Local Universe.

Still concerning the SDSS blue galaxies sample, with redshifts $0 < z < 0.30$, analyzed in this work, in future work we intend to study the universe dynamics through the analysis of the multipole correlation function~\citep{Anderson14} and the application of reconstruction methods~\citep{Burden14}.

The reconstruction tool is an interesting approach that is being applied in recent BAO analyses with the main goal of increasing the statistical significance of the BAO-scale measurement and decreasing the error estimation, while the position of the BAO scale remains, basically, the same. 
The disadvantage in obtaining this result is the need to employ a fiducial cosmology, including a non-linear clustering model on small scales. 
Ultimately, the statistically significant BAO measurement is model dependent (see, e.g.,~\cite{Beutler11} and 
\cite{Carter18} to compare BAO measurements without and with reconstruction procedure, respectively). 
On the other hand, as described in section~\ref{sec:Introduction}, our choice is to analyse the SDSS blue galaxies, that show reduced effects of non-linear clustering because they are found in low density regions, 
making it possible to fit the 2PCF without assuming a cosmological model, allowing us to perform a BAO-scale measurement at low redshift with minimal model assumptions. 



\section*{Acknowledgements} 
The authors thank the referee for useful comments and suggestions, which significantly increased the quality of our manuscript. 
FA thanks the Conselho Nacional de Desenvolvimento Cient\'{i}fico e Tecnologico (CNPq, National Council for Scientific and Technological Development) and the Fundação Carlos Chagas Filho de Amparo à Pesquisa do Estado do Rio de Janeiro, 
FAPERJ - Processo SEI 260003/014913/2023, for their financial support. RCN thanks the financial support from the CNPq for partial financial support under the project No. 304306/2022-3, and the Fundação de Amparo à pesquisa do Estado do RS (FAPERGS, Research Support Foundation of the State of RS) for partial financial support under the project No. 23/2551-0000848-3. AB acknowledges a CNPq fellowship.

\section*{Data Availability}

The data underlying this article were accessed from \url{https://data.sdss.org/sas/dr12/}.
The derived data generated in this research will be shared on reasonable request to the corresponding author.



\bibliographystyle{mnras}
\bibliography{bao} 

\begin{thebibliography}{}
\makeatletter
\relax
\def\mn@urlcharsother{\let\do\@makeother \do\$\do\&\do\#\do\^\do\_\do\%\do\~}
\def\mn@doi{\begingroup\mn@urlcharsother \@ifnextchar [ {\mn@doi@}
  {\mn@doi@[]}}
\def\mn@doi@[#1]#2{\def\@tempa{#1}\ifx\@tempa\@empty \href
  {http://dx.doi.org/#2} {doi:#2}\else \href {http://dx.doi.org/#2} {#1}\fi
  \endgroup}
\def\mn@eprint#1#2{\mn@eprint@#1:#2::\@nil}
\def\mn@eprint@arXiv#1{\href {http://arxiv.org/abs/#1} {{\tt arXiv:#1}}}
\def\mn@eprint@dblp#1{\href {http://dblp.uni-trier.de/rec/bibtex/#1.xml}
  {dblp:#1}}
\def\mn@eprint@#1:#2:#3:#4\@nil{\def\@tempa {#1}\def\@tempb {#2}\def\@tempc
  {#3}\ifx \@tempc \@empty \let \@tempc \@tempb \let \@tempb \@tempa \fi \ifx
  \@tempb \@empty \def\@tempb {arXiv}\fi \@ifundefined
  {mn@eprint@\@tempb}{\@tempb:\@tempc}{\expandafter \expandafter \csname
  mn@eprint@\@tempb\endcsname \expandafter{\@tempc}}}

\bibitem[\protect\citeauthoryear{{Abbott} et~al.,}{{Abbott}
  et~al.}{2019}]{Abbott19}
{Abbott} T.~M.~C.,  et~al., 2019, \mn@doi [\mnras] {10.1093/mnras/sty3351},
  \href {https://ui.adsabs.harvard.edu/abs/2019MNRAS.483.4866A} {483, 4866}

\bibitem[\protect\citeauthoryear{{Agrawal}, {Makiya}, {Chiang}, {Jeong},
  {Saito}  \& {Komatsu}}{{Agrawal} et~al.}{2017}]{Agrawal17}
{Agrawal} A.,  {Makiya} R.,  {Chiang} C.-T.,  {Jeong} D.,  {Saito} S.,
  {Komatsu} E.,  2017, \mn@doi [\jcap] {10.1088/1475-7516/2017/10/003}, \href
  {https://ui.adsabs.harvard.edu/abs/2017JCAP...10..003A} {2017, 003}

\bibitem[\protect\citeauthoryear{{Akarsu}, {Di Valentino}, {Kumar}, {Nunes},
  {Vazquez}  \& {Yadav}}{{Akarsu} et~al.}{2023}]{Akarsu23}
{Akarsu} O.,  {Di Valentino} E.,  {Kumar} S.,  {Nunes} R.~C.,  {Vazquez} J.~A.,
    {Yadav} A.,  2023, \mn@doi [arXiv e-prints] {10.48550/arXiv.2307.10899},
  \href {https://ui.adsabs.harvard.edu/abs/2023arXiv230710899A} {p.
  arXiv:2307.10899}

\bibitem[\protect\citeauthoryear{{Alam} et~al.,}{{Alam} et~al.}{2021}]{Alam21}
{Alam} S.,  et~al., 2021, \mn@doi [\prd] {10.1103/PhysRevD.103.083533}, \href
  {https://ui.adsabs.harvard.edu/abs/2021PhRvD.103h3533A} {103, 083533}

\bibitem[\protect\citeauthoryear{{Anderson} et~al.,}{{Anderson}
  et~al.}{2014}]{Anderson14}
{Anderson} L.,  et~al., 2014, \mn@doi [\mnras] {10.1093/mnras/stu523}, \href
  {https://ui.adsabs.harvard.edu/abs/2014MNRAS.441...24A} {441, 24}

\bibitem[\protect\citeauthoryear{{Anselmi}, {Starkman}  \& {Renzi}}{{Anselmi}
  et~al.}{2023}]{Anselmi22}
{Anselmi} S.,  {Starkman} G.~D.,   {Renzi} A.,  2023, \mn@doi [\prd]
  {10.1103/PhysRevD.107.123506}, \href
  {https://ui.adsabs.harvard.edu/abs/2023PhRvD.107l3506A} {107, 123506}

\bibitem[\protect\citeauthoryear{{Avila}, {Novaes}, {Bernui}  \& {de
  Carvalho}}{{Avila} et~al.}{2018}]{Avila18}
{Avila} F.,  {Novaes} C.~P.,  {Bernui} A.,   {de Carvalho} E.,  2018, \mn@doi
  [\jcap] {10.1088/1475-7516/2018/12/041}, \href
  {https://ui.adsabs.harvard.edu/abs/2018JCAP...12..041A} {12, 041}

\bibitem[\protect\citeauthoryear{{Avila}, {Novaes}, {Bernui}, {de Carvalho}  \&
  {Nogueira-Cavalcante}}{{Avila} et~al.}{2019}]{Avila19}
{Avila} F.,  {Novaes} C.~P.,  {Bernui} A.,  {de Carvalho} E.,
  {Nogueira-Cavalcante} J.~P.,  2019, \mn@doi [\mnras] {10.1093/mnras/stz1765},
  \href {https://ui.adsabs.harvard.edu/abs/2019MNRAS.488.1481A} {488, 1481}

\bibitem[\protect\citeauthoryear{{Avila}, {Bernui}, {de Carvalho}  \&
  {Novaes}}{{Avila} et~al.}{2021}]{Avila21}
{Avila} F.,  {Bernui} A.,  {de Carvalho} E.,   {Novaes} C.~P.,  2021, \mn@doi
  [\mnras] {10.1093/mnras/stab1488}, \href
  {https://ui.adsabs.harvard.edu/abs/2021MNRAS.505.3404A} {505, 3404}

\bibitem[\protect\citeauthoryear{{Avila}, {Bernui}, {Nunes}, {de Carvalho}  \&
  {Novaes}}{{Avila} et~al.}{2022}]{Avila22}
{Avila} F.,  {Bernui} A.,  {Nunes} R.~C.,  {de Carvalho} E.,   {Novaes} C.~P.,
  2022, \mn@doi [\mnras] {10.1093/mnras/stab3122}, \href
  {https://ui.adsabs.harvard.edu/abs/2022MNRAS.509.2994A} {509, 2994}

\bibitem[\protect\citeauthoryear{{Avila}, {Oliveira}, {Dias}  \&
  {Bernui}}{{Avila} et~al.}{2023}]{Avila23}
{Avila} F.,  {Oliveira} J.,  {Dias} M. L.~S.,   {Bernui} A.,  2023, \mn@doi
  [Brazilian Journal of Physics] {10.1007/s13538-023-01259-z}, \href
  {https://ui.adsabs.harvard.edu/abs/2023BrJPh..53...49A} {53, 49}

\bibitem[\protect\citeauthoryear{{Bassett} \& {Hlozek}}{{Bassett} \&
  {Hlozek}}{2010}]{Bassett10}
{Bassett} B.,  {Hlozek} R.,  2010, in {Ruiz-Lapuente} P.,  ed., , Dark Energy:
  Observational and Theoretical Approaches.
Cambridge University Press Cambridge, p.~246, \mn@doi{10.48550/arXiv.0910.5224}

\bibitem[\protect\citeauthoryear{{Benisty} \& {Staicova}}{{Benisty} \&
  {Staicova}}{2021}]{Benisty21}
{Benisty} D.,  {Staicova} D.,  2021, \mn@doi [\aap]
  {10.1051/0004-6361/202039502}, \href
  {https://ui.adsabs.harvard.edu/abs/2021A&A...647A..38B} {647, A38}

\bibitem[\protect\citeauthoryear{{Benisty}, {Mifsud}, {Levi Said}  \&
  {Staicova}}{{Benisty} et~al.}{2023}]{Benisty23}
{Benisty} D.,  {Mifsud} J.,  {Levi Said} J.,   {Staicova} D.,  2023, \mn@doi
  [Physics of the Dark Universe] {10.1016/j.dark.2023.101344}, \href
  {https://ui.adsabs.harvard.edu/abs/2023PDU....4201344B} {42, 101344}

\bibitem[\protect\citeauthoryear{{Beutler} et~al.,}{{Beutler}
  et~al.}{2011}]{Beutler11}
{Beutler} F.,  et~al., 2011, \mn@doi [\mnras]
  {10.1111/j.1365-2966.2011.19250.x}, \href
  {https://ui.adsabs.harvard.edu/abs/2011MNRAS.416.3017B} {416, 3017}

\bibitem[\protect\citeauthoryear{Beutler et~al.,}{Beutler
  et~al.}{2016}]{Beutler_2016}
Beutler F.,  et~al., 2016, \mn@doi [Monthly Notices of the Royal Astronomical
  Society] {10.1093/mnras/stw2373}, 464, 3409–3430

\bibitem[\protect\citeauthoryear{{Blake} et~al.,}{{Blake}
  et~al.}{2011}]{Blake11}
{Blake} C.,  et~al., 2011, \mn@doi [\mnras] {10.1111/j.1365-2966.2011.19592.x},
  \href {https://ui.adsabs.harvard.edu/abs/2011MNRAS.418.1707B} {418, 1707}

\bibitem[\protect\citeauthoryear{{Blot} et~al.,}{{Blot} et~al.}{2019}]{Blot19}
{Blot} L.,  et~al., 2019, \mn@doi [\mnras] {10.1093/mnras/stz507}, \href
  {https://ui.adsabs.harvard.edu/abs/2019MNRAS.485.2806B} {485, 2806}

\bibitem[\protect\citeauthoryear{{Bond} \& {Efstathiou}}{{Bond} \&
  {Efstathiou}}{1984}]{Bond84}
{Bond} J.~R.,  {Efstathiou} G.,  1984, \mn@doi [\apjl] {10.1086/184362}, \href
  {https://ui.adsabs.harvard.edu/abs/1984ApJ...285L..45B} {285, L45}

\bibitem[\protect\citeauthoryear{{Burden}, {Percival}, {Manera}, {Cuesta},
  {Vargas Magana}  \& {Ho}}{{Burden} et~al.}{2014}]{Burden14}
{Burden} A.,  {Percival} W.~J.,  {Manera} M.,  {Cuesta} A.~J.,  {Vargas Magana}
  M.,   {Ho} S.,  2014, \mn@doi [\mnras] {10.1093/mnras/stu1965}, \href
  {https://ui.adsabs.harvard.edu/abs/2014MNRAS.445.3152B} {445, 3152}

\bibitem[\protect\citeauthoryear{{Carnero}, {S{\'a}nchez}, {Crocce},
  {Cabr{\'e}}  \& {Gazta{\~n}aga}}{{Carnero} et~al.}{2012}]{Carnero12}
{Carnero} A.,  {S{\'a}nchez} E.,  {Crocce} M.,  {Cabr{\'e}} A.,
  {Gazta{\~n}aga} E.,  2012, \mn@doi [\mnras]
  {10.1111/j.1365-2966.2011.19832.x}, \href
  {https://ui.adsabs.harvard.edu/abs/2012MNRAS.419.1689C} {419, 1689}

\bibitem[\protect\citeauthoryear{{Carter}, {Beutler}, {Percival}, {Blake},
  {Koda}  \& {Ross}}{{Carter} et~al.}{2018}]{Carter18}
{Carter} P.,  {Beutler} F.,  {Percival} W.~J.,  {Blake} C.,  {Koda} J.,
  {Ross} A.~J.,  2018, \mn@doi [\mnras] {10.1093/mnras/sty2405}, \href
  {https://ui.adsabs.harvard.edu/abs/2018MNRAS.481.2371C} {481, 2371}

\bibitem[\protect\citeauthoryear{{Carter}, {Beutler}, {Percival}, {DeRose},
  {Wechsler}  \& {Zhao}}{{Carter} et~al.}{2020}]{Carter20}
{Carter} P.,  {Beutler} F.,  {Percival} W.~J.,  {DeRose} J.,  {Wechsler} R.~H.,
    {Zhao} C.,  2020, \mn@doi [\mnras] {10.1093/mnras/staa761}, \href
  {https://ui.adsabs.harvard.edu/abs/2020MNRAS.494.2076C} {494, 2076}

\bibitem[\protect\citeauthoryear{{Chilingarian} \& {Zolotukhin}}{{Chilingarian}
  \& {Zolotukhin}}{2012}]{Chilingarian12}
{Chilingarian} I.~V.,  {Zolotukhin} I.~Y.,  2012, \mn@doi [\mnras]
  {10.1111/j.1365-2966.2011.19837.x}, \href
  {https://ui.adsabs.harvard.edu/abs/2012MNRAS.419.1727C} {419, 1727}

\bibitem[\protect\citeauthoryear{{Chilingarian}, {Melchior}  \&
  {Zolotukhin}}{{Chilingarian} et~al.}{2010}]{Chilingarian10}
{Chilingarian} I.~V.,  {Melchior} A.-L.,   {Zolotukhin} I.~Y.,  2010, \mn@doi
  [\mnras] {10.1111/j.1365-2966.2010.16506.x}, \href
  {https://ui.adsabs.harvard.edu/abs/2010MNRAS.405.1409C} {405, 1409}

\bibitem[\protect\citeauthoryear{{Colavincenzo} et~al.,}{{Colavincenzo}
  et~al.}{2019}]{Colavincenzo19}
{Colavincenzo} M.,  et~al., 2019, \mn@doi [\mnras] {10.1093/mnras/sty2964},
  \href {https://ui.adsabs.harvard.edu/abs/2019MNRAS.482.4883C} {482, 4883}

\bibitem[\protect\citeauthoryear{{Cole} et~al.,}{{Cole} et~al.}{2005}]{Cole05}
{Cole} S.,  et~al., 2005, \mn@doi [\mnras] {10.1111/j.1365-2966.2005.09318.x},
  \href {https://ui.adsabs.harvard.edu/abs/2005MNRAS.362..505C} {362, 505}

\bibitem[\protect\citeauthoryear{{Coles} \& {Jones}}{{Coles} \&
  {Jones}}{1991}]{Coles91}
{Coles} P.,  {Jones} B.,  1991, \mn@doi [\mnras] {10.1093/mnras/248.1.1}, \href
  {https://ui.adsabs.harvard.edu/abs/1991MNRAS.248....1C} {248, 1}

\bibitem[\protect\citeauthoryear{{Courtois}, {Pomar{\`e}de}, {Tully}, {Hoffman}
   \& {Courtois}}{{Courtois} et~al.}{2013}]{Courtois13}
{Courtois} H.~M.,  {Pomar{\`e}de} D.,  {Tully} R.~B.,  {Hoffman} Y.,
  {Courtois} D.,  2013, \mn@doi [\aj] {10.1088/0004-6256/146/3/69}, \href
  {https://ui.adsabs.harvard.edu/abs/2013AJ....146...69C} {146, 69}

\bibitem[\protect\citeauthoryear{{Crocce} \& {Scoccimarro}}{{Crocce} \&
  {Scoccimarro}}{2008}]{Crocce08}
{Crocce} M.,  {Scoccimarro} R.,  2008, \mn@doi [\prd]
  {10.1103/PhysRevD.77.023533}, \href
  {https://ui.adsabs.harvard.edu/abs/2008PhRvD..77b3533C} {77, 023533}

\bibitem[\protect\citeauthoryear{{Croton}, {Norberg}, {Gazta{\~n}aga}  \&
  {Baugh}}{{Croton} et~al.}{2007}]{Croton07}
{Croton} D.~J.,  {Norberg} P.,  {Gazta{\~n}aga} E.,   {Baugh} C.~M.,  2007,
  \mn@doi [\mnras] {10.1111/j.1365-2966.2007.12035.x}, \href
  {https://ui.adsabs.harvard.edu/abs/2007MNRAS.379.1562C} {379, 1562}

\bibitem[\protect\citeauthoryear{{D'Agostino} \& {Nunes}}{{D'Agostino} \&
  {Nunes}}{2023}]{DAgostino23}
{D'Agostino} R.,  {Nunes} R.~C.,  2023, \mn@doi [\prd]
  {10.1103/PhysRevD.108.023523}, \href
  {https://ui.adsabs.harvard.edu/abs/2023PhRvD.108b3523D} {108, 023523}

\bibitem[\protect\citeauthoryear{{Dias}, {Avila}  \& {Bernui}}{{Dias}
  et~al.}{2023}]{Dias23}
{Dias} B.~L.,  {Avila} F.,   {Bernui} A.,  2023, \mn@doi [\mnras]
  {10.1093/mnras/stad2980}, \href
  {https://ui.adsabs.harvard.edu/abs/2023MNRAS.526.3219D} {526, 3219}

\bibitem[\protect\citeauthoryear{{Dinda}}{{Dinda}}{2023}]{Bikash23}
{Dinda} B.~R.,  2023, \mn@doi [arXiv e-prints] {10.48550/arXiv.2311.13498},
  \href {https://ui.adsabs.harvard.edu/abs/2023arXiv231113498D} {p.
  arXiv:2311.13498}

\bibitem[\protect\citeauthoryear{{Eisenstein} \& {Hu}}{{Eisenstein} \&
  {Hu}}{1998}]{Einsenstein98}
{Eisenstein} D.~J.,  {Hu} W.,  1998, \mn@doi [\apj] {10.1086/305424}, \href
  {https://ui.adsabs.harvard.edu/abs/1998ApJ...496..605E} {496, 605}

\bibitem[\protect\citeauthoryear{{Eisenstein} et~al.,}{{Eisenstein}
  et~al.}{2005}]{Eisenstein05}
{Eisenstein} D.~J.,  et~al., 2005, \mn@doi [\apj] {10.1086/466512}, \href
  {https://ui.adsabs.harvard.edu/abs/2005ApJ...633..560E} {633, 560}

\bibitem[\protect\citeauthoryear{{Feldman}, {Kaiser}  \& {Peacock}}{{Feldman}
  et~al.}{1994}]{Feldman94}
{Feldman} H.~A.,  {Kaiser} N.,   {Peacock} J.~A.,  1994, \mn@doi [\apj]
  {10.1086/174036}, \href
  {https://ui.adsabs.harvard.edu/abs/1994ApJ...426...23F} {426, 23}

\bibitem[\protect\citeauthoryear{{Foreman-Mackey}, {Hogg}, {Lang}  \&
  {Goodman}}{{Foreman-Mackey} et~al.}{2013}]{Foreman-Mackey_2013}
{Foreman-Mackey} D.,  {Hogg} D.~W.,  {Lang} D.,   {Goodman} J.,  2013, \mn@doi
  [\pasp] {10.1086/670067}, \href
  {https://ui.adsabs.harvard.edu/abs/2013PASP..125..306F} {125, 306}

\bibitem[\protect\citeauthoryear{{Franco}, {Avila}  \& {Bernui}}{{Franco}
  et~al.}{2024}]{Franco24}
{Franco} C.,  {Avila} F.,   {Bernui} A.,  2024, \mn@doi [\mnras]
  {10.1093/mnras/stad3616}, \href
  {https://ui.adsabs.harvard.edu/abs/2024MNRAS.527.7400F} {527, 7400}

\bibitem[\protect\citeauthoryear{{Gerke} et~al.,}{{Gerke}
  et~al.}{2007}]{Gerke07}
{Gerke} B.~F.,  et~al., 2007, \mn@doi [\mnras]
  {10.1111/j.1365-2966.2007.11374.x}, \href
  {https://ui.adsabs.harvard.edu/abs/2007MNRAS.376.1425G} {376, 1425}

\bibitem[\protect\citeauthoryear{{Giar{\`e}}, {G{\'o}mez-Valent}, {Di
  Valentino}  \& {van de Bruck}}{{Giar{\`e}} et~al.}{2024}]{Giare23}
{Giar{\`e}} W.,  {G{\'o}mez-Valent} A.,  {Di Valentino} E.,   {van de Bruck}
  C.,  2024, \mn@doi [\prd] {10.1103/PhysRevD.109.063516}, \href
  {https://ui.adsabs.harvard.edu/abs/2024PhRvD.109f3516G} {109, 063516}

\bibitem[\protect\citeauthoryear{{Hand}, {Feng}, {Beutler}, {Li}, {Modi},
  {Seljak}  \& {Slepian}}{{Hand} et~al.}{2018}]{Hand18}
{Hand} N.,  {Feng} Y.,  {Beutler} F.,  {Li} Y.,  {Modi} C.,  {Seljak} U.,
  {Slepian} Z.,  2018, \mn@doi [\aj] {10.3847/1538-3881/aadae0}, \href
  {https://ui.adsabs.harvard.edu/abs/2018AJ....156..160H} {156, 160}

\bibitem[\protect\citeauthoryear{{Harris} et~al.,}{{Harris}
  et~al.}{2020}]{Harris20}
{Harris} C.~R.,  et~al., 2020, \mn@doi [\nat] {10.1038/s41586-020-2649-2},
  \href {https://ui.adsabs.harvard.edu/abs/2020Natur.585..357H} {585, 357}

\bibitem[\protect\citeauthoryear{{He}, {Zhao}  \& {Shan}}{{He}
  et~al.}{2023}]{He23}
{He} M.,  {Zhao} C.,   {Shan} H.,  2023, \mn@doi [\mnras]
  {10.1093/mnras/stad2207}, \href
  {https://ui.adsabs.harvard.edu/abs/2023MNRAS.525.1746H} {525, 1746}

\bibitem[\protect\citeauthoryear{Heinesen, Blake, Li  \& Wiltshire}{Heinesen
  et~al.}{2019}]{Heinesen_2019}
Heinesen A.,  Blake C.,  Li Y.-Z.,   Wiltshire D.~L.,  2019, \mn@doi [Journal
  of Cosmology and Astroparticle Physics] {10.1088/1475-7516/2019/03/003},
  2019, 003–003

\bibitem[\protect\citeauthoryear{{Hoffman}, {Pomar{\`e}de}, {Tully}  \&
  {Courtois}}{{Hoffman} et~al.}{2017}]{Hoffman17}
{Hoffman} Y.,  {Pomar{\`e}de} D.,  {Tully} R.~B.,   {Courtois} H.~M.,  2017,
  \mn@doi [Nature Astronomy] {10.1038/s41550-016-0036}, \href
  {https://ui.adsabs.harvard.edu/abs/2017NatAs...1E..36H} {1, 0036}

\bibitem[\protect\citeauthoryear{{Jarvis}, {Bernstein}  \& {Jain}}{{Jarvis}
  et~al.}{2004}]{Jarvis04}
{Jarvis} M.,  {Bernstein} G.,   {Jain} B.,  2004, \mn@doi [\mnras]
  {10.1111/j.1365-2966.2004.07926.x}, \href
  {https://ui.adsabs.harvard.edu/abs/2004MNRAS.352..338J} {352, 338}

\bibitem[\protect\citeauthoryear{{Jones} et~al.,}{{Jones}
  et~al.}{2004}]{Jones04}
{Jones} D.~H.,  et~al., 2004, \mn@doi [\mnras]
  {10.1111/j.1365-2966.2004.08353.x}, \href
  {https://ui.adsabs.harvard.edu/abs/2004MNRAS.355..747J} {355, 747}

\bibitem[\protect\citeauthoryear{{Keih{\"a}nen} et~al.,}{{Keih{\"a}nen}
  et~al.}{2019}]{Keihanen19}
{Keih{\"a}nen} E.,  et~al., 2019, \mn@doi [\aap] {10.1051/0004-6361/201935828},
  \href {https://ui.adsabs.harvard.edu/abs/2019A&A...631A..73K} {631, A73}

\bibitem[\protect\citeauthoryear{{Kerscher}, {Szapudi}  \& {Szalay}}{{Kerscher}
  et~al.}{2000}]{Kerscher00}
{Kerscher} M.,  {Szapudi} I.,   {Szalay} A.~S.,  2000, \mn@doi [\apjl]
  {10.1086/312702}, \href
  {https://ui.adsabs.harvard.edu/abs/2000ApJ...535L..13K} {535, L13}

\bibitem[\protect\citeauthoryear{{Kester}, {Bernui}  \&
  {Hip{\'o}lito-Ricaldi}}{{Kester} et~al.}{2024}]{Kester23}
{Kester} C.~E.,  {Bernui} A.,   {Hip{\'o}lito-Ricaldi} W.~S.,  2024, \mn@doi
  [\aap] {10.1051/0004-6361/202348160}, \href
  {https://ui.adsabs.harvard.edu/abs/2024A&A...683A.176K} {683, A176}

\bibitem[\protect\citeauthoryear{{Landy} \& {Szalay}}{{Landy} \&
  {Szalay}}{1993}]{Landy93}
{Landy} S.~D.,  {Szalay} A.~S.,  1993, \mn@doi [\apj] {10.1086/172900}, \href
  {https://ui.adsabs.harvard.edu/abs/1993ApJ...412...64L} {412, 64}

\bibitem[\protect\citeauthoryear{{Lewis} \& {Challinor}}{{Lewis} \&
  {Challinor}}{2011}]{CAMB}
{Lewis} A.,  {Challinor} A.,  2011, {CAMB: Code for Anisotropies in the
  Microwave Background}, Astrophysics Source Code Library, record ascl:1102.026
  (\mn@eprint {ascl} {1102.026})

\bibitem[\protect\citeauthoryear{{Li}, {Du}  \& {Xu}}{{Li} et~al.}{2020}]{Li20}
{Li} E.-K.,  {Du} M.,   {Xu} L.,  2020, \mn@doi [\mnras]
  {10.1093/mnras/stz3308}, \href
  {https://ui.adsabs.harvard.edu/abs/2020MNRAS.491.4960L} {491, 4960}

\bibitem[\protect\citeauthoryear{{Lippich} et~al.,}{{Lippich}
  et~al.}{2019}]{Lippich19}
{Lippich} M.,  et~al., 2019, \mn@doi [\mnras] {10.1093/mnras/sty2757}, \href
  {https://ui.adsabs.harvard.edu/abs/2019MNRAS.482.1786L} {482, 1786}

\bibitem[\protect\citeauthoryear{{Liu}, {Qiao}, {Chang}  \& {Xu}}{{Liu}
  et~al.}{2023}]{Liu23}
{Liu} J.,  {Qiao} L.,  {Chang} B.,   {Xu} L.,  2023, \mn@doi [European Physical
  Journal C] {10.1140/epjc/s10052-023-11545-4}, \href
  {https://ui.adsabs.harvard.edu/abs/2023EPJC...83..374L} {83, 374}

\bibitem[\protect\citeauthoryear{{Marques} \& {Bernui}}{{Marques} \&
  {Bernui}}{2020}]{Marques20}
{Marques} G.~A.,  {Bernui} A.,  2020, \mn@doi [\jcap]
  {10.1088/1475-7516/2020/05/052}, \href
  {https://ui.adsabs.harvard.edu/abs/2020JCAP...05..052M} {05, 052}

\bibitem[\protect\citeauthoryear{{Marques}, {Novaes}, {Bernui}  \&
  {Ferreira}}{{Marques} et~al.}{2018}]{Marques18}
{Marques} G.~A.,  {Novaes} C.~P.,  {Bernui} A.,   {Ferreira} I.~S.,  2018,
  \mn@doi [\mnras] {10.1093/mnras/stx2240}, \href
  {https://ui.adsabs.harvard.edu/abs/2018MNRAS.473..165M} {473, 165}

\bibitem[\protect\citeauthoryear{Marra \& Isidro}{Marra \&
  Isidro}{2019}]{Marra_2019}
Marra V.,  Isidro E. G.~C.,  2019, \mn@doi [\mnras] {10.1093/mnras/stz1557},
  487, 3419

\bibitem[\protect\citeauthoryear{{Marulli}, {Veropalumbo}  \&
  {Moresco}}{{Marulli} et~al.}{2016}]{Marulli16}
{Marulli} F.,  {Veropalumbo} A.,   {Moresco} M.,  2016, \mn@doi [Astronomy and
  Computing] {10.1016/j.ascom.2016.01.005}, \href
  {https://ui.adsabs.harvard.edu/abs/2016A&C....14...35M} {14, 35}

\bibitem[\protect\citeauthoryear{{Mohammad} et~al.,}{{Mohammad}
  et~al.}{2018}]{Mohammad18}
{Mohammad} F.~G.,  et~al., 2018, \mn@doi [\aap] {10.1051/0004-6361/201731685},
  \href {https://ui.adsabs.harvard.edu/abs/2018A&A...610A..59M} {610, A59}

\bibitem[\protect\citeauthoryear{{Peebles} \& {Hauser}}{{Peebles} \&
  {Hauser}}{1974}]{Peebles74}
{Peebles} P.~J.~E.,  {Hauser} M.~G.,  1974, \mn@doi [\apjs] {10.1086/190308},
  \href {https://ui.adsabs.harvard.edu/abs/1974ApJS...28...19P} {28, 19}

\bibitem[\protect\citeauthoryear{{Peebles} \& {Yu}}{{Peebles} \&
  {Yu}}{1970}]{Peebles70}
{Peebles} P.~J.~E.,  {Yu} J.~T.,  1970, \mn@doi [\apj] {10.1086/150713}, \href
  {https://ui.adsabs.harvard.edu/abs/1970ApJ...162..815P} {162, 815}

\bibitem[\protect\citeauthoryear{{Percival} et~al.,}{{Percival}
  et~al.}{2010}]{Percival10}
{Percival} W.~J.,  et~al., 2010, \mn@doi [\mnras]
  {10.1111/j.1365-2966.2009.15812.x}, \href
  {https://ui.adsabs.harvard.edu/abs/2010MNRAS.401.2148P} {401, 2148}

\bibitem[\protect\citeauthoryear{{Planck Collaboration} et~al.,}{{Planck
  Collaboration} et~al.}{2020}]{Planck20}
{Planck Collaboration} et~al., 2020, \mn@doi [\aap]
  {10.1051/0004-6361/201833910}, \href
  {https://ui.adsabs.harvard.edu/abs/2020A&A...641A...6P} {641, A6}

\bibitem[\protect\citeauthoryear{{Ram{\'\i}rez-P{\'e}rez}, {Sanchez}, {Alonso}
  \& {Font-Ribera}}{{Ram{\'\i}rez-P{\'e}rez} et~al.}{2022}]{Ramirez22}
{Ram{\'\i}rez-P{\'e}rez} C.,  {Sanchez} J.,  {Alonso} D.,   {Font-Ribera} A.,
  2022, \mn@doi [\jcap] {10.1088/1475-7516/2022/05/002}, \href
  {https://ui.adsabs.harvard.edu/abs/2022JCAP...05..002R} {2022, 002}

\bibitem[\protect\citeauthoryear{{Ross} et~al.,}{{Ross} et~al.}{2014}]{Ross14}
{Ross} A.~J.,  et~al., 2014, \mn@doi [\mnras] {10.1093/mnras/stt1895}, \href
  {https://ui.adsabs.harvard.edu/abs/2014MNRAS.437.1109R} {437, 1109}

\bibitem[\protect\citeauthoryear{{S{\'a}nchez} et~al.,}{{S{\'a}nchez}
  et~al.}{2011}]{Sanchez11}
{S{\'a}nchez} E.,  et~al., 2011, \mn@doi [\mnras]
  {10.1111/j.1365-2966.2010.17679.x}, \href
  {https://ui.adsabs.harvard.edu/abs/2011MNRAS.411..277S} {411, 277}

\bibitem[\protect\citeauthoryear{{Schlegel}, {Finkbeiner}  \&
  {Davis}}{{Schlegel} et~al.}{1998}]{Schlegel98}
{Schlegel} D.~J.,  {Finkbeiner} D.~P.,   {Davis} M.,  1998, \mn@doi [\apj]
  {10.1086/305772}, \href
  {https://ui.adsabs.harvard.edu/abs/1998ApJ...500..525S} {500, 525}

\bibitem[\protect\citeauthoryear{{Springel} et~al.,}{{Springel}
  et~al.}{2005}]{Springel05}
{Springel} V.,  et~al., 2005, \mn@doi [\nat] {10.1038/nature03597}, \href
  {https://ui.adsabs.harvard.edu/abs/2005Natur.435..629S} {435, 629}

\bibitem[\protect\citeauthoryear{{Staicova} \& {Benisty}}{{Staicova} \&
  {Benisty}}{2022}]{Staicova22}
{Staicova} D.,  {Benisty} D.,  2022, \mn@doi [\aap]
  {10.1051/0004-6361/202244366}, \href
  {https://ui.adsabs.harvard.edu/abs/2022A&A...668A.135S} {668, A135}

\bibitem[\protect\citeauthoryear{{Sunyaev} \& {Zeldovich}}{{Sunyaev} \&
  {Zeldovich}}{1970}]{Sunyaev70}
{Sunyaev} R.~A.,  {Zeldovich} Y.~B.,  1970, \mn@doi [\apss]
  {10.1007/BF00653471}, \href
  {https://ui.adsabs.harvard.edu/abs/1970Ap&SS...7....3S} {7, 3}

\bibitem[\protect\citeauthoryear{{Tegmark}}{{Tegmark}}{1997}]{Tegmark97}
{Tegmark} M.,  1997, \mn@doi [\prl] {10.1103/PhysRevLett.79.3806}, \href
  {https://ui.adsabs.harvard.edu/abs/1997PhRvL..79.3806T} {79, 3806}

\bibitem[\protect\citeauthoryear{{Tully}, {Pomar{\`e}de}, {Graziani},
  {Courtois}, {Hoffman}  \& {Shaya}}{{Tully} et~al.}{2019}]{Tully19}
{Tully} R.~B.,  {Pomar{\`e}de} D.,  {Graziani} R.,  {Courtois} H.~M.,
  {Hoffman} Y.,   {Shaya} E.~J.,  2019, \mn@doi [\apj]
  {10.3847/1538-4357/ab2597}, \href
  {https://ui.adsabs.harvard.edu/abs/2019ApJ...880...24T} {880, 24}

\bibitem[\protect\citeauthoryear{Tully, Howlett  \& Pomarede}{Tully
  et~al.}{2023}]{tully2023hooleilana}
Tully R.~B.,  Howlett C.,   Pomarede D.,  2023, Ho'oleilana: An Individual
  Baryon Acoustic Oscillation? (\mn@eprint {arXiv} {2309.00677})

\bibitem[\protect\citeauthoryear{{Vargas-Maga{\~n}a}
  et~al.,}{{Vargas-Maga{\~n}a} et~al.}{2013}]{Vargas13}
{Vargas-Maga{\~n}a} M.,  et~al., 2013, \mn@doi [\aap]
  {10.1051/0004-6361/201220790}, \href
  {https://ui.adsabs.harvard.edu/abs/2013A&A...554A.131V} {554, A131}

\bibitem[\protect\citeauthoryear{{Visser}}{{Visser}}{2005}]{Visser05}
{Visser} M.,  2005, \mn@doi [General Relativity and Gravitation]
  {10.1007/s10714-005-0134-8}, \href
  {https://ui.adsabs.harvard.edu/abs/2005GReGr..37.1541V} {37, 1541}

\bibitem[\protect\citeauthoryear{{Vogelsberger} et~al.,}{{Vogelsberger}
  et~al.}{2014}]{Vogelsberger14}
{Vogelsberger} M.,  et~al., 2014, \mn@doi [\mnras] {10.1093/mnras/stu1536},
  \href {https://ui.adsabs.harvard.edu/abs/2014MNRAS.444.1518V} {444, 1518}

\bibitem[\protect\citeauthoryear{{Weinberg}, {Mortonson}, {Eisenstein},
  {Hirata}, {Riess}  \& {Rozo}}{{Weinberg} et~al.}{2013}]{Weinberg13}
{Weinberg} D.~H.,  {Mortonson} M.~J.,  {Eisenstein} D.~J.,  {Hirata} C.,
  {Riess} A.~G.,   {Rozo} E.,  2013, \mn@doi [\physrep]
  {10.1016/j.physrep.2013.05.001}, \href
  {https://ui.adsabs.harvard.edu/abs/2013PhR...530...87W} {530, 87}

\bibitem[\protect\citeauthoryear{{Xavier}, {Abdalla}  \& {Joachimi}}{{Xavier}
  et~al.}{2016}]{Xavier16}
{Xavier} H.~S.,  {Abdalla} F.~B.,   {Joachimi} B.,  2016, \mn@doi [\mnras]
  {10.1093/mnras/stw874}, \href
  {https://ui.adsabs.harvard.edu/abs/2016MNRAS.459.3693X} {459, 3693}

\bibitem[\protect\citeauthoryear{{York} et~al.,}{{York} et~al.}{2000}]{York00}
{York} D.~G.,  et~al., 2000, \mn@doi [\aj] {10.1086/301513}, \href
  {https://ui.adsabs.harvard.edu/abs/2000AJ....120.1579Y} {120, 1579}

\bibitem[\protect\citeauthoryear{{Zehavi} et~al.,}{{Zehavi}
  et~al.}{2005}]{Zehavi05}
{Zehavi} I.,  et~al., 2005, \mn@doi [\apj] {10.1086/431891}, \href
  {https://ui.adsabs.harvard.edu/abs/2005ApJ...630....1Z} {630, 1}

\bibitem[\protect\citeauthoryear{{de Carvalho}, {Bernui}, {Carvalho}, {Novaes}
  \& {Xavier}}{{de Carvalho} et~al.}{2018}]{Edilson18}
{de Carvalho} E.,  {Bernui} A.,  {Carvalho} G.~C.,  {Novaes} C.~P.,   {Xavier}
  H.~S.,  2018, \mn@doi [\jcap] {10.1088/1475-7516/2018/04/064}, \href
  {https://ui.adsabs.harvard.edu/abs/2018JCAP...04..064D} {04, 064}

\bibitem[\protect\citeauthoryear{{de Carvalho}, {Bernui}, {Xavier}  \&
  {Novaes}}{{de Carvalho} et~al.}{2020}]{Edilson20}
{de Carvalho} E.,  {Bernui} A.,  {Xavier} H.~S.,   {Novaes} C.~P.,  2020,
  \mn@doi [\mnras] {10.1093/mnras/staa119}, \href
  {https://ui.adsabs.harvard.edu/abs/2020MNRAS.492.4469D} {492, 4469}

\bibitem[\protect\citeauthoryear{{de Carvalho}, {Bernui}, {Avila}, {Novaes}  \&
  {Nogueira-Cavalcante}}{{de Carvalho} et~al.}{2021}]{Edilson21}
{de Carvalho} E.,  {Bernui} A.,  {Avila} F.,  {Novaes} C.~P.,
  {Nogueira-Cavalcante} J.~P.,  2021, \mn@doi [\aap]
  {10.1051/0004-6361/202039936}, \href
  {https://ui.adsabs.harvard.edu/abs/2021A&A...649A..20D} {649, A20}

\makeatother
\end{thebibliography}

\appendix
\section{Triangle posteriors and tables}
\label{appendix}

\begin{table*}
\renewcommand{\arraystretch}{1.5}
    \centering
    \caption{Summary information of the best-fit at 1$\sigma$ CL for all baseline parameters.}
    \label{tab:bao_fit_ps}
\scalebox{1.1}{
    \begin{tabular}{ccccc}
        \hline
		Parameters    & Sample 1 & Sample 2 & Sample 3 & Sample 4\\ 
		\hline
  
		A             & $0.03 \pm 0.01$        & $0.02 \pm 0.01$        &   $0.03\pm 0.01$      &   $0.03 \pm 0.01$     \\ 
            B         & $445.46^{+102}_{-101}$        & $470.81^{87.47}_{98.81}$        &   $339.24^{108.40}_{84.13}$      &   $363.55^{120.60}_{88.01}$     \\  
		  $\delta$      & $-1.01 \pm 0.01$        & $-1.01 \pm 0.01$        &   $-1.01 \pm 0.01$      &   $-1.01 \pm 0.01$     \\  
		C             & $0.01^{+0.01}_{-0.009}$        & $0.01^{0.01}_{0.009}$        &   $0.01^{+0.01}_{-0.009}$      &   $0.01^{+0.01}_{-0.009}$     \\  
            D             & $-433.40^{101.55}_{102.29}$        & $-458.23^{98.81}_{87.47}$        &   $-327.91^{84.18}_{108.30}$      &   $-352.07^{87.95}_{120.67}$     \\  
            $\Sigma$      & $40.73^{38.05}_{15.99}$        & $54.51^{32.48}_{23.35}$        &   $35.53^{29.94}_{12.19}$      &   $35.43^{32.40}_{12.21}$     \\ 
        $s_{\text{BAO}}$  & $100.28^{10.79}_{22.96}$        & $108.94^{13.69}_{16.70}$        &   $87.28^{9.26}_{21.86}$      &   $89.21^{9.24}_{23.02}$     \\ 
		\hline
    \end{tabular}
}
\end{table*}


\begin{table*}
\renewcommand{\arraystretch}{1.5}
    \centering
    \caption{Summary information of the best-fit at 1$\sigma$ CL for all baseline parameters, assuming $\alpha$ as a free parameter.}
    \label{tab:bao_fit_ps_alfa_free}
\scalebox{1.1}{
    \begin{tabular}{ccccc}
        \hline
		Parameters    & Sample 1 & Sample 2 & Sample 3 & Sample 4\\ 
		\hline
  
		A             & $0.03 \pm 0.01$        & $0.02 \pm 0.01$        &   $0.03\pm 0.01$      &   $0.03 \pm 0.01$     \\ 
            B         & $460.10^{+91.32}_{-100.71}$        & $481.82^{+80.37}_{-92.75}$        &   $347.61^{+102.05}_{-74.93}$      &   $379.52^{+106.79}_{-85.72}$     \\  
		  $\delta$      & $-1.01 \pm 0.01$        & $-1.01 \pm 0.01$        &   $-1.01 \pm 0.01$      &   $-1.01 \pm 0.01$     \\  
		C             & $0.01^{+0.02}_{-0.009}$        & $0.01^{+0.02}_{-0.009}$        &   $0.01^{+0.01}_{-0.009}$      &   $0.01^{+0.01}_{-0.009}$     \\  
            D             & $-448.11^{+100.74}_{-91.14}$        & $-469.23^{+92.66}_{-80.37}$        &   $-336.28^{+74.98}_{-102.18}$      &   $-368.16^{+85.73}_{-106.61}$     \\  
            $\Sigma$      & $42.08^{+37.40}_{-19.98}$        & $50.40^{+32.88}_{-24.08}$        &   $37.21^{+32.83}_{-16.38}$      &   $34.88^{+30.95}_{-14.65}$     \\ 
        $s_{\text{BAO}}$  & $88.92^{+38.06}_{-32.59}$        & $92.76^{+35.49}_{-30.78}$        &   $80.87^{+35.64}_{-29.64}$      &   $84.49^{+35.06}_{-30.60}$     \\ 
        $\alpha$ & $0.91^{+0.36}_{-0.29}$ & $0.87^{+0.33}_{-0.26}$ & $1.01^{+0.33}_{-0.33}$ & $0.99^{+0.35}_{-0.33}$ \\
		\hline
    \end{tabular}
}
\end{table*}

\begin{figure*}
         \centering
        \includegraphics[width=\textwidth]{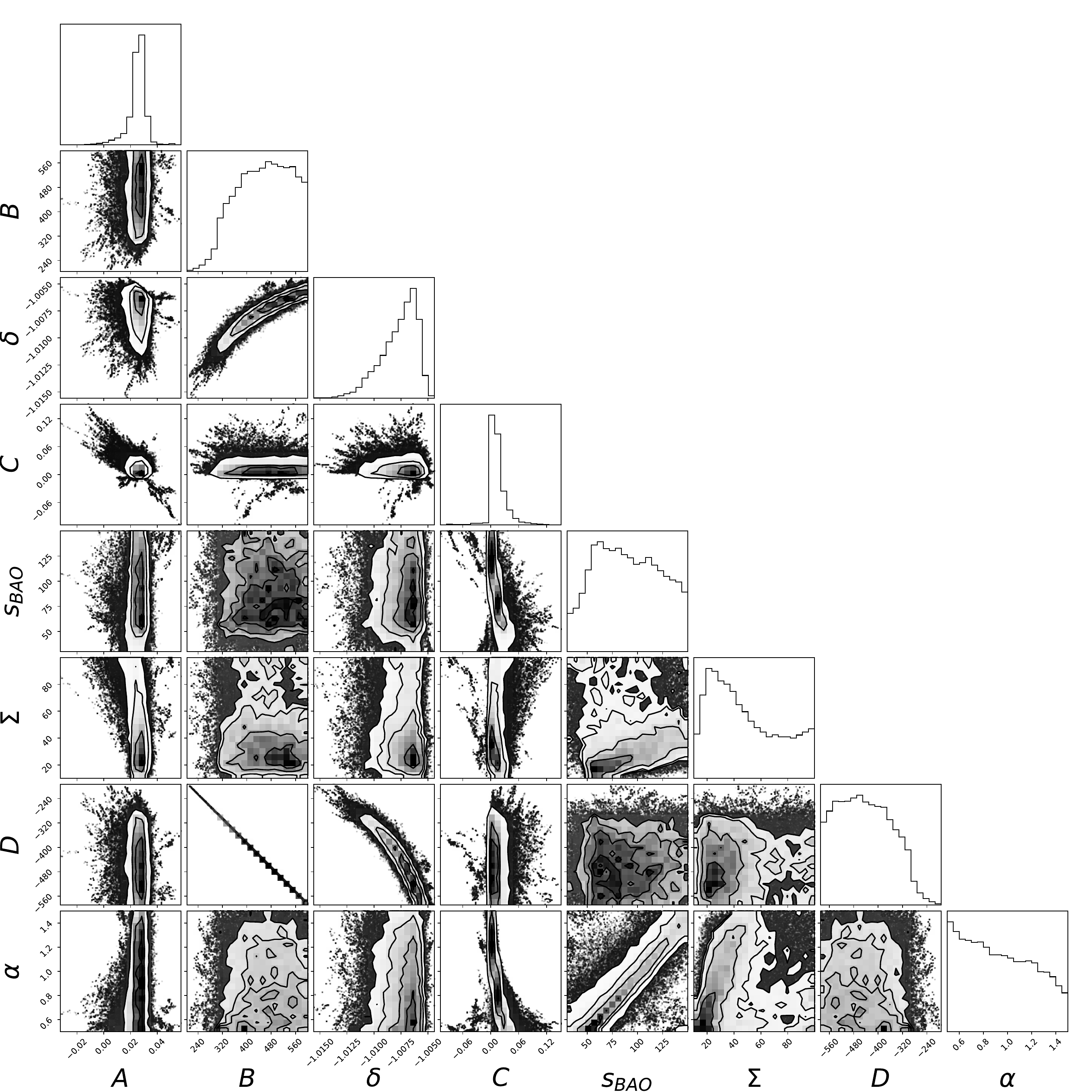}
        \caption{Depicts the 1- and 2-dimensional marginalised distributions of all free parameters employed in our fitting process for Sample 1.}
         \label{fig:1a}
\end{figure*}

\begin{figure*}
         \centering
        \includegraphics[width=\textwidth]{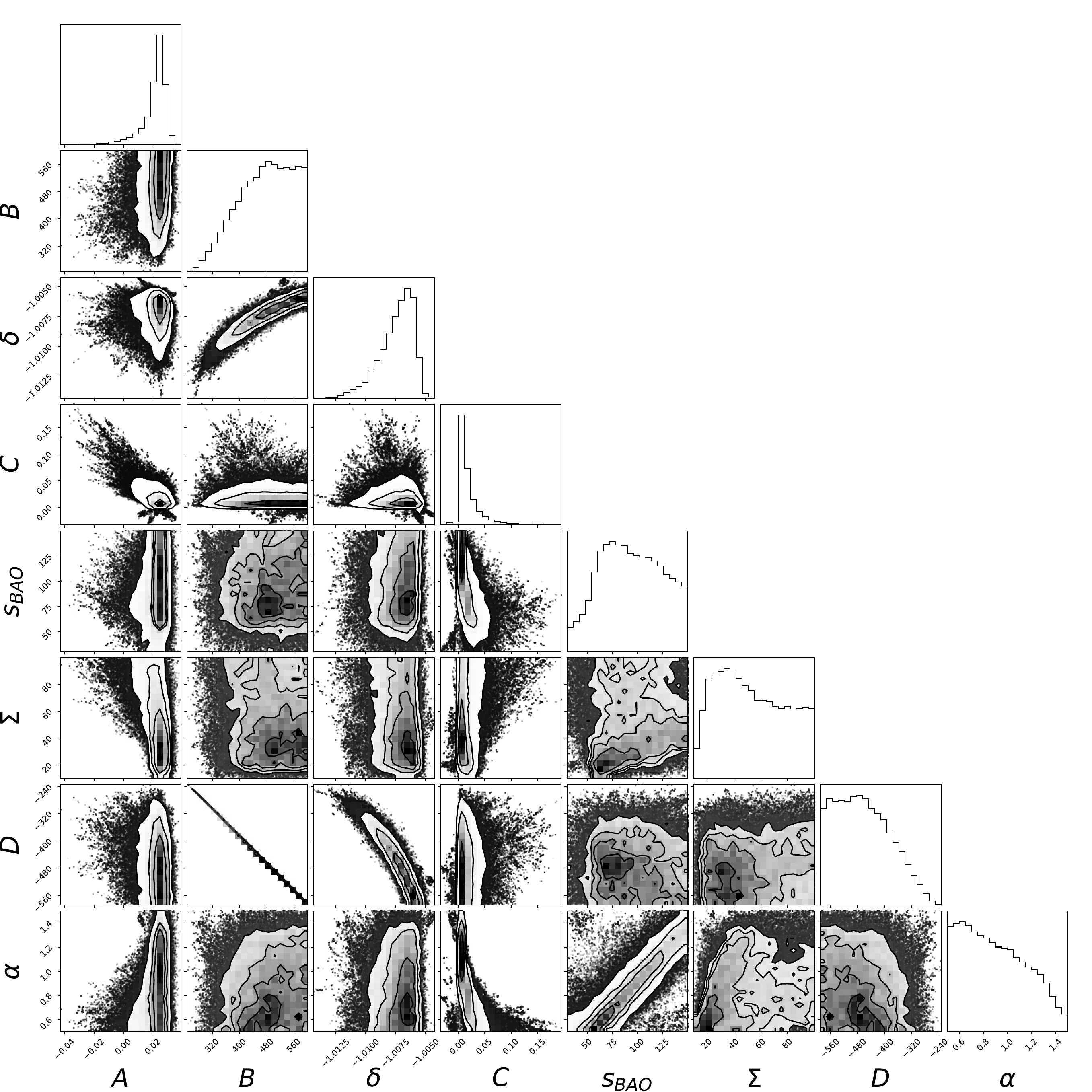}
        \caption{Same as in figure \ref{fig:1a}, but for Sample 2.}
         \label{fig:2a}
\end{figure*}

\begin{figure*}
         \centering
        \includegraphics[width=\textwidth]{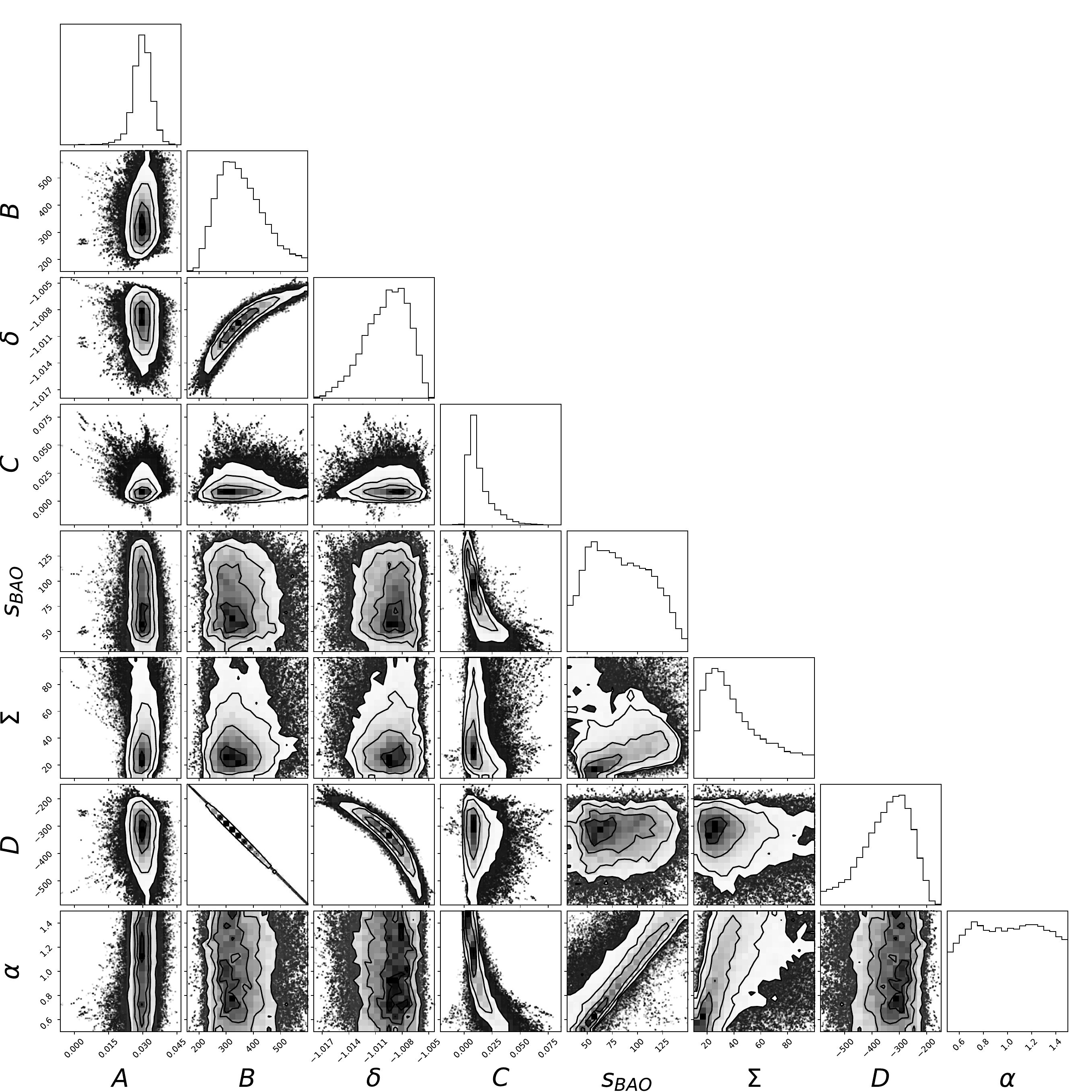}
        \caption{Same as in figure \ref{fig:1a}, but for Sample 3.}
         \label{fig:3a}
\end{figure*}

\begin{figure*}
         \centering
        \includegraphics[width=\textwidth]{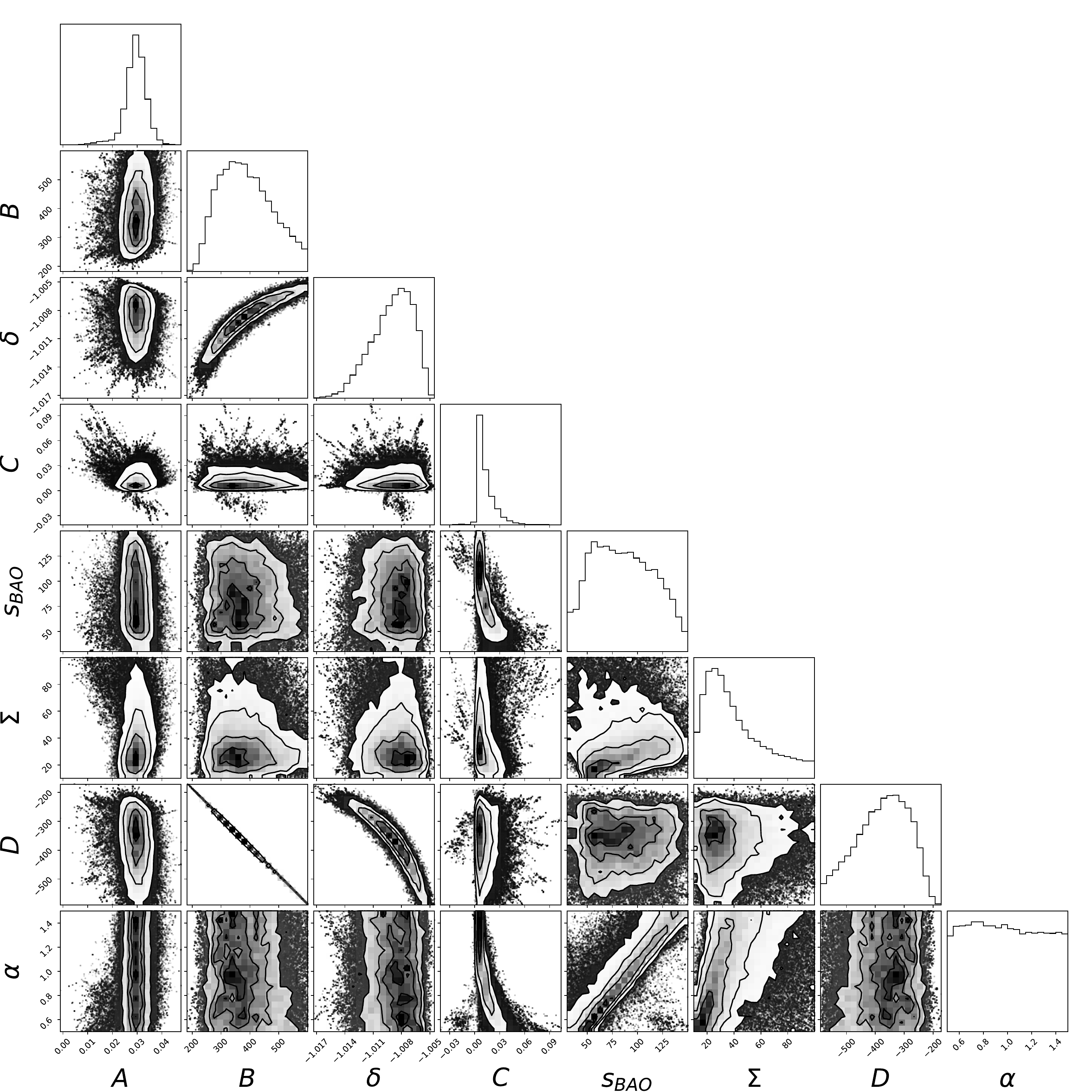}
        \caption{Same as in figure \ref{fig:1a}, but for Sample 4.}
         \label{fig:4a}
\end{figure*}

In Figures \ref{fig:1a}, \ref{fig:2a},  \ref{fig:3a} and \ref{fig:4a} we showcase the 1- and 2-dimensional projections of the posterior probability distributions across the entire parametric space for the four samples investigated in this study.

Tables \ref{tab:bao_fit_ps} and \ref{tab:bao_fit_ps_alfa_free} provides a comprehensive summary of the statistical analyses conducted on the main parameters considered in our fit of the Two-Point Correlation Function (2PCF), as depicted in Figures \ref{fig:best_fit_models_1_2} and \ref{fig:best_fit_models_3_4}.

\bsp	
\label{lastpage}
\end{document}